\documentclass[11pt]{article}

\usepackage[final]{acl}

\usepackage{times}
\usepackage{latexsym}

\usepackage[T1]{fontenc}

\usepackage[utf8]{inputenc}

\usepackage{microtype}

\usepackage{inconsolata}

\usepackage{graphicx}

\usepackage{times}
\usepackage{latexsym}
\usepackage{booktabs}
\usepackage{amssymb}
\usepackage{amsmath}
\usepackage{fixltx2e}
\usepackage[most]{tcolorbox} 
\usepackage{listings} 
\usepackage{listings}
\usepackage{mdframed} 
\definecolor{verylightgray}{gray}{0.97}

\usepackage{pifont}

\usepackage{tcolorbox}
\usepackage[table]{xcolor}

\newtcolorbox{promptbox}[1][]{
  colback=verylightgray, 
  colframe=lightgray, 
  title=Prompt, 
  fonttitle=\small\bfseries, 
  #1,
}

\definecolor{backgroundcolor}{rgb}{0.97, 0.97, 0.97}
\definecolor{delim}{RGB}{20,105,176}
\colorlet{punct}{red!60!black}
\definecolor{numb}{rgb}{0.5,0,0.5}
\definecolor{darkgray}{rgb}{0.4, 0.4, 0.4}
\definecolor{keywordcolor}{rgb}{0.4, 0.4, 0.95}

\lstdefinelanguage{json}{
    basicstyle=\scriptsize\ttfamily,
    frame=single,
    rulecolor=\color{darkgray}, 
    backgroundcolor=\color{backgroundcolor},
    showstringspaces=false,  
    showspaces=false,
    columns=flexible,
    morekeywords={user_history, generated_queries},
    keywordstyle=\color{keywordcolor}\bfseries
}

\usepackage[T1]{fontenc}

\usepackage[utf8]{inputenc}
\usepackage{graphicx}
\usepackage{multirow}
\usepackage{subcaption}

\usepackage{microtype}
\usepackage{algorithm}
\usepackage{algpseudocode}
\usepackage{inconsolata}
\usepackage{amsmath} 

\usepackage{graphicx}

\title{Rethinking LLM-Based Recommendations: \\A Personalized Query-Driven Parallel Integration}

\author{
Donghee Han~~~~~~ Hwanjun Song$^{\dagger}$~~~~~~ Mun Yong Yi$^{\dagger}$ \smallskip \\
KAIST, Republic of Korea\\  
\texttt{\{handonghee,  songhwanjun, munyi\}@kaist.ac.kr}\\
\smallskip
\small{
\textsuperscript{\dag}Corresponding authors
}
}

\newcommand{\algname}{\textsc{QueRec}}

\newcommand{\perfmax}{57\%}
\newcommand{\perfavg}{31\%}
\newcommand{\fix}[1]{{#1}}

\begin{document}
\maketitle
\begin{abstract}
Recent studies have explored integrating large language models (LLMs) into recommendation systems but face several challenges, including training-induced bias and bottlenecks from serialized architecture.
To effectively address these issues, we propose a \textbf{Query-to-Recommendation}, a \textbf{parallel recommendation framework} that decouples LLMs from candidate pre-selection and instead enables direct retrieval over the entire item pool. 
Our framework \textbf{connects LLMs and recommendation models} in a parallel manner, allowing each component to independently utilize its strengths without interfering with the other.
In this framework, LLMs are utilized to generate feature-enriched item descriptions and personalized user queries, allowing for capturing diverse preferences and enabling rich semantic matching in a zero-shot manner.
To effectively combine the complementary strengths of LLM and collaborative signals, we introduce an adaptive reranking strategy.
Extensive experiments demonstrate an improvement in performance up to \perfmax{}, while also improving the novelty and diversity of recommendations.

%
\end{abstract}

\section{Introduction}
Recommendation systems play a crucial role in delivering personalized content and service across various domains. As the demand for more accurate and diverse recommendations continues to increase, the integration of large language models (LLM) has emerged as a promising advancement~\cite{llmrec_survey_0,llmrec_survey_1}. 
LLMs possess extensive knowledge and exhibit remarkable abilities in understanding and generating text~\cite{llm_survey}, enabling new opportunities to improve recommendation systems beyond the traditional collaborative filtering (CF) and content-based methods. Recently, numerous studies have been proposed to leverage LLM in recommendation systems~\cite{llm_user_profile,llm_user_pref,llm_control}.

\fix{
To harness the capabilities of LLMs for recommendation, two major research paradigms have emerged.
The first line of works utilize LLMs as generative predictors, typically fine-tuning them on next-item prediction tasks. 
These methods generate textual representations (e.g., item titles) based on user histories and use them as queries for retrieving candidate items~\cite{bigrec, gpt4rec}. 
While this approach leverages LLMs’ generation strength, it requires fine-tuning, which introduces bias to train dataset and reduces diversity of recommendation. 
Additionally, relying solely on item titles can limit the expressive capacity of LLMs, limiting their ability to align nuanced user preferences with rich item characteristics.
}

\fix{
The second line of research focuses on reranking candidate items directly within the LLM prompt. 
In this setting, a separate candidate retrieval model selects a subset of items, which are then presented to the LLM for scoring or reranking~\cite{llmrank, allmrec, tallrec}. 
These approaches typically leverage CF-based recommendation models such as SASRec~\cite{sasrec} for candidate selection~\cite{palr}, and various techniques have been proposed to align LLMs with CF signals~\cite{allmrec,screc}.

However, these approaches inherently depend on the performance of the candidate selector, and they limit the ability of LLMs to fully leverage their distinctive capabilities across the entire item pool. 
The information handled by the CF-based model and the LLM is fundamentally different, and the current serialized architecture can limit performance due to the misalignment between these two components.
This misalignment also becomes a bottleneck that hinders the utilization of the diverse knowledge embedded in LLMs for recommendation.
}


\begin{figure*}[t]
    \centering
    \includegraphics[width=0.85\linewidth]{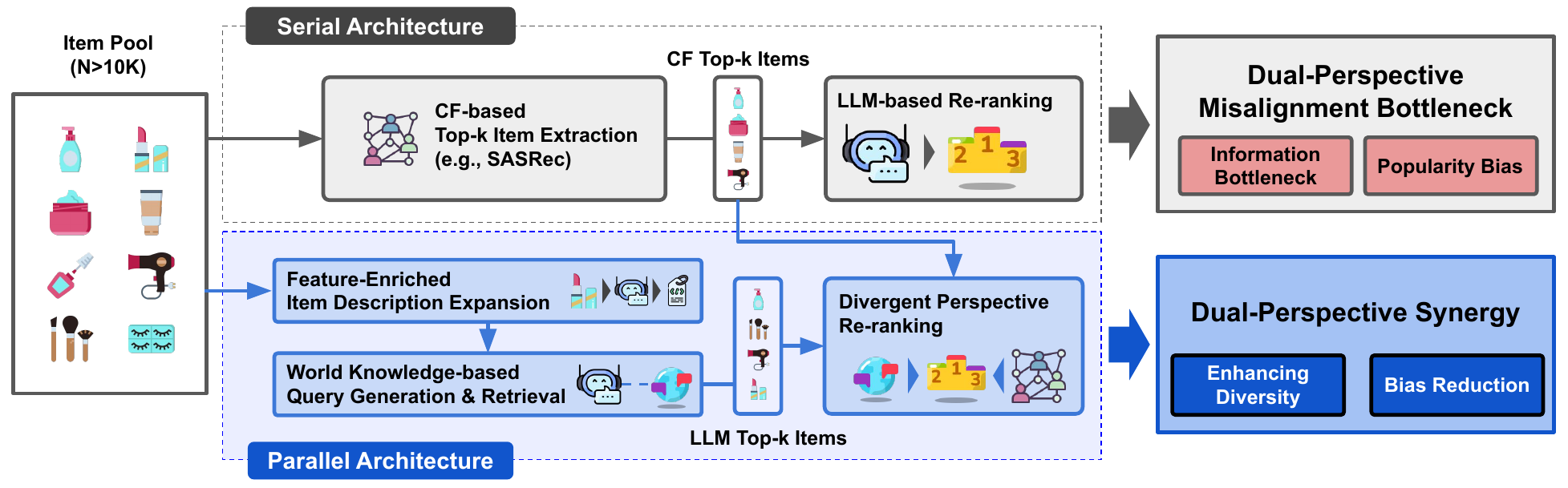}
    \vspace*{-0.3cm}
    \caption{Comparison between the traditional serialized pipeline and the proposed parallel approach.}
    \label{fig:motiv}
    \vspace*{-0.4cm}
\end{figure*}

\fix{
In this study, we focus on addressing two key challenges in LLM-based recommendation: the bias introduced by fine-tuning LLMs on recommendation datasets, and the bottleneck caused by the serialized architecture that utilizes CF-based models as candidate selectors. 
Motivated by the fact that CF-based models and LLMs handle fundamentally different types of information, we propose a parallel framework that integrates both components without entangling their processes. 
Our approach enables the generation of enriched, feature-aware queries that allow LLMs to effectively leverage their broad and diverse knowledge for recommendation.
}

Fig.\ref{fig:motiv} illustrates the difference between the traditional serialized approach and our proposed parallel approach. Our proposed parallel architecture maximizes the utilization of the LLM’s world knowledge and creates synergy by effectively merging the two complementary perspectives.
We argue that LLMs can offer a complementary perspective to CF-based recommendation models. 
To fully leverage their potential, LLMs need direct access to the global item pool; however, existing approaches that fine-tune LLMs to generate item titles hinder the utilization of the diverse knowledge of LLMs and introduce biases from the training dataset..

\fix{In our framework, LLMs and CF-based models independently retrieve top-k items from the entire candidate pool. 
By decoupling LLMs from pre-selected candidates, our method maximizes LLMs’ ability and creates synergy between complementary perspectives.
To utilize LLMs' rich world knowledge, we introduce item description expansion and personalized query generation where the LLM generates feature-enriched item description and personalized user queries grounded in user histories and item characteristics.
%
Furthermore, we design an adaptive reranking strategy that aggregates the outputs of LLMs and CF-based models while maintaining their independency. 
Unlike existing approaches that jointly train multiple models, our method dynamically balances contributions.
}

Our method not only improves the performance of the recommendation, but also promotes greater diversity and adaptability, effectively addressing the limitations of existing LLM-based recommendation systems. 
The training-free nature of our framework allows for seamless incorporation of new items based on their key features without additional model updates, and it can also handle large, dynamically changing item pools. 
This advantage is particularly beneficial in real-world e-commerce scenarios, where large volumes of new items are continuously introduced and updated. 
Consequently, our design significantly reduces both infrastructure costs and maintenance overhead, making the system highly scalable for rapidly evolving recommendation environments.

\begin{figure*}[!ht]
    \centering
    \vspace*{-3mm}
    \includegraphics[width=0.92\textwidth]{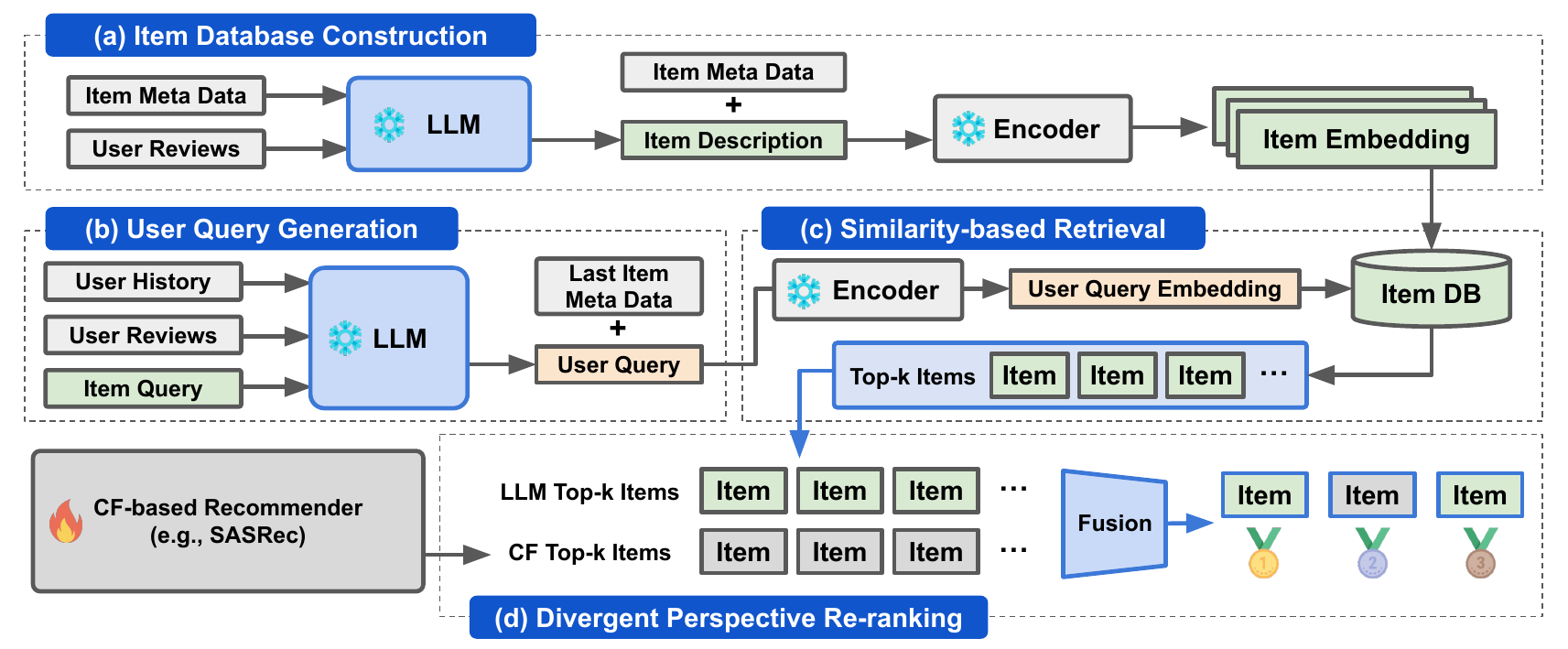}
    \vspace*{-0.2cm}
    \caption{Overview of \algname{}.  }
    \label{fig:main}
    \vspace*{-3mm}
\end{figure*}

Our contributions in this work are as follows.

\begin{itemize}
    \item \fix{We \textbf{introduce a parallel architecture} for recommendation that combines LLM-based retrieval and CF-based models, maximizing the complementary strengths of each component.}
    \item Our method \textbf{outperforms state-of-the-art methods} and achieved up to \perfmax{} improvement in performance over traditional CF-based models, without requiring additional LLM fine-tuning.
    \item \fix{Through our analysis, we demonstrate that our approach effectively \textbf{addresses bias and diversity issues} in recommendations.}
\end{itemize}

\section{Related Work}
\subsection{Traditional Recommendation Systems}
Sequential recommendation has emerged as a prominent paradigm, leveraging users' recent interactions to predict the next likely item of interest. 
Deep learning-based models, such as SASRec~\cite{sasrec} and BERT4Rec~\cite{bert4rec}, have demonstrated strong performance in capturing complex sequential user behaviors, effectively modeling temporal dependencies to improve the precision of the recommendation. 
Recommendation systems that rely solely on item IDs fail to capture the key features of the items. To address this limitation, various sequential recommendation models have been proposed that incorporate additional information \cite{sasrecd,morec}.




\subsection{LLM-based Recommendation Systems}
With the rapid advancement of LLMs, there has been a growing interest in incorporating them into recommendation systems \cite{llmrec_survey_0}. 
Early LLM-based recommendation approaches fine-tuned pre-trained models such as T5~\cite{t5} to generate item recommendations by providing item or user IDs and output item IDs in a text format \cite{p5,pod, rdrec, elmrec}. However, these approaches had limitations in fully leveraging the rich knowledge embedded in LLMs. 

Another direction in LLM-based recommendation research explores natural language prompts instead of item and user IDs, allowing LLMs to leverage their extensive knowledge and zero-shot capabilities. 
Early studies fine-tuned LLMs \cite{tallrec}, while later approaches fixed LLM parameters and update only additional components to incorporate CF-signals \cite{allmrec,screc}. 
Recent work introduced zero-shot LLM-based reranking~\citet{llmrank}, but these methods remain limited by the reliance on upstream candidate selectors and positional bias. 

Recently, other studies proposed fine-tuning LLMs to generate item titles and using the generated titles as search queries for recommendation\cite{gpt4rec,bigrec}.
However, these approaches have limitations in leveraging the diverse knowledge inherent in LLMs.
To address these challenges, we propose a method that better aligns with real-world recommendation settings, ensuring ease of deployment and scalability in large-scale item pools.




\section{Method: \algname{} Framework}
In this section, we introduce \algname (Query-to-Recommendation), a novel framework designed to leverage the LLMs through large item pool and to operate in parallel with CF-based models, enabling a dual-perspective that integrates both collaborative signals and language-based knowledge.
Fig.\ref{fig:main} illustrates the overall process of our method.
The proposed method leverages LLMs to generate feature-enriched item descriptions and personalized user queries, retrieving top-k items based on similarity scores between user queries and item descriptions.
These LLM-based results are then combined with the top-k items from a CF-based model, effectively fusing insights from both CF-based models and LLMs.


\subsection{Item Database Construction}
To effectively leverage the world knowledge of LLMs to highlight key features of items and align them with user preferences, we employ LLMs to generate feature-enriched item descriptions that reflect the distinctive features of each item.
Fig.\ref{fig:main}(a) illustrates the process of leveraging an LLM to expand item descriptions, which are then encoded into embeddings to construct an item vector database.

This process involves constructing prompts using item metadata (e.g. title, category, description) and user reviews to provide contextual information to the LLM. 
We design the prompt to expand item descriptions into a set of diverse queries, aiming to better reflect users’ varied preferences. Given these prompts, the LLM generates ten distinct queries, each capturing different aspects of the item by leveraging its extensive knowledge.
Finally, the generated queries are combined with the original item metadata to create a comprehensive item representation, which is then converted into an item embedding using a retrieval encoder and cached for search.

\vspace{-0.5cm}
\begin{equation}
    q_i = \text{LLM}(p_i, m_i, r_i), \quad
    d_i = m_i \oplus q_i
\label{equ:item_query}
\end{equation}
\vspace{-0.5cm}

Equation (\ref{equ:item_query}) describes the process of extracting item queries and constructing the complete item representation corpus. Given an item $i$, we construct a query $q_i$ using the LLM with inputs: item query generation prompt $p_i$, item metadata $m_i$, and user reviews $r_i$. The final textual representation $d_i$ is formed by concatenating the metadata of the item with the generated query. Here, $\oplus$ represents the concatenation operator, integrating both structured metadata and insights derived from LLM to improve the retrieval effectiveness of items.

\vspace{-0.3cm}
\begin{equation}
\begin{aligned}
\mathcal{V} = \{\text{Enc}(d_i) \mid d_i \in \mathcal{D} \}
\end{aligned}
\label{equ:item_db}
\end{equation}

\noindent Continuing from the previous step, we encode all enriched item descriptions $d_i \in \mathcal{D}$ using a retrieval encoder and cache the resulting embeddings $\mathcal{V}$ for efficient search during recommendation, as shown in Equation (\ref{equ:item_db}).

\subsection{User Query Generation}
To represent the user, we incorporate information from their historical interactions and preferences. 
This process begins with the construction of prompts that combine user history (including item titles and metadata), user reviews, and previously generated item descriptions to provide rich contextual information. 
Based on these prompts, the LLM generates ten personalized queries that reflect the user's context and preferences. 
To emphasize the user’s most recent preferences, we append the enriched description of the most recently purchased item to the prompt.

\vspace{-0.5cm}
\begin{equation}
q_u = \text{LLM}(p_u, h_u, r_u, d_l), \quad
d_u = m_l \oplus q_u 
\label{equ:user_query}
\end{equation}
\vspace{-0.5cm}

Equation (\ref{equ:user_query}) describes the process of constructing the textual user representation. Given a user $u$, we construct a query $q_u$ using the LLM with inputs: user query generation prompt $p_u$, user history $h_u$ (which includes item titles and metadata), user reviews $r_u$, and the most recent item description $d_l$. The final user representation $d_u$ is obtained by concatenating the user history with the generated query and incorporating the last interacted item metadata $m_l$ to better reflect temporal user preferences. 

An important aspect of this step is that, instead of using each generated query independently, we aggregate them into a single unified user representation. 
Similar to the item description expansion, we prompt the LLM to generate multiple personalized queries in order to capture diverse user perspectives and preferences. 
These queries are not treated separately but are concatenated into a single user representation, which is then used to construct a comprehensive user embedding.
The prompt format and generated query examples are described in the Appendix~\ref{apd_prompt}.

\subsection{Similarity-based Retrieval}
With the user textual representations and item database obtained from the previous steps, we perform similarity-based retrieval for personalized recommendation :

\vspace{-0.2cm}
\begin{equation}
\begin{aligned}
v_u &= \text{Enc}(d_u), \quad v_i \in \mathcal{V} \\
s_{u,i} &= \cos(v_u, v_i)
\end{aligned}
\label{equ:enc}
\end{equation}
\vspace{-0.4cm}

\vspace{-0.5cm}
\begin{equation}
\hat{\mathcal{I}}_u = \underset{i \in \mathcal{I}}{\text{arg max } s_{u,i}}, \quad \text{where } |\hat{\mathcal{I}}_u| = k
\label{equ:topk}
\end{equation}
\vspace{-0.5cm}

\noindent Equation (\ref{equ:enc}) defines the embedding extraction via the pre-trained text encoder $\text{Enc}$ and similarity computation between the user embedding $v_u$ and item embeddings in item database $v_i \in \mathcal{V}$. We compute cosine similarity $s_{u,i}$ between the embeddings to measure their relevance. Based on these scores, we select the top-k items $\hat{\mathcal{I}}_u$ for the user query, as shown in Equation (\ref{equ:topk}), which are the most relevant items to the user's preferences.

Our semantic similarity-based retrieval utilizing personalized queries offers richer information compared to traditional grounding methods that rely solely on item titles.
It is also more adaptable to updates in the item pool, which occur frequently in real-world recommendation scenarios.

\subsection{\fix{Divergent Perspective Reranking}}

To effectively integrate LLM-based semantic insights and CF-based collaborative signals, we propose a divergent perspective reranking method. 
Initially, similarity scores from each model (LLM and CF) are normalized, ensuring fair influence from each method:

\begin{equation}
\vspace{-1mm}
\begin{aligned}
&\tilde{s}_{u,i}^{X} = 
\frac{s_{u,i}^{X} - \min(s_{u}^{X})}
{\max(s_{u}^{X}) - \min(s_{u}^{X})}, \\
&X \in \{LLM, CF\} .
\label{equ:score_norm}
\end{aligned}
\end{equation}

\noindent In Equation (\ref{equ:score_norm}), $s_{u,i}^{X}$ and $\tilde{s}_{u,i}^{X}$ denote the original score and final normalized score, between user $u$ and item $i$ computed by model $X$ (LLM-based and CF-based). 
For each user, we apply min-max normalization over the score set $s_u^X = \{ s_{u,i}^X \mid i \in \mathcal{I}_u \}$, where $\mathcal{I}_u$ represents the entire item pool. 
This normalization maps all user-item relevance scores to the range $[0, 1]$, ensuring that the scores from different models are comparable on the same scale. 
The process preserves the internal ranking structure of each model while aligning their score magnitudes for reranking.

Subsequently, we calculate the final reranking score using a convex combination (CC) of the normalized scores:

\begin{equation}
\vspace{-1mm}
s_{u,i}^{*} = \lambda \tilde{s}_{u,i}^{LLM} + (1 - \lambda) \tilde{s}_{u,i}^{CF} .
\label{equ:final_score}
\end{equation}

\noindent The initial $\lambda$ is determined by the validation Hit@10 performance of each model:

\begin{equation}
\vspace{-1mm}
\lambda_{\text{init}} = \frac{H_{10}^{LLM}}{H_{10}^{LLM} + H_{10}^{CF}} .
\label{equ:lambda}
\end{equation}

\noindent
This method of computing the initial $\lambda$ is intended to assign a larger weight to the model with higher performance. However, it may be less effective in cases where the two models perform well on entirely different user groups.

Further analysis revealed that the intersection ratio between hit items from \algname{} and CF models varies significantly across datasets and models. In the case of a low intersection ratio with imbalanced weights (e.g., 0.3), interference between the two recommended item lists causes performance to decrease. To address this, we adjust $\lambda$ by incorporating the intersection ratio:

\begin{equation}
\vspace{-2mm}
\lambda = \omega \cdot \lambda_{\text{init}} + (1 - \omega) \cdot 0.5 .
\end{equation}

\noindent Here, $\omega$ is computed as:

\begin{equation}
\vspace{-1mm}
\omega = \frac{|\mathcal{H}_{10}^{\text{LLM}} \cap \mathcal{H}_{10}^{\text{CF}}|}{|\mathcal{H}_{10}^{\text{LLM}} \cup \mathcal{H}_{10}^{\text{CF}}|} ,
\end{equation}

\noindent where $\mathcal{H}_{10}^{\text{LLM}}$ and $\mathcal{H}_{10}^{\text{CF}}$ denote the sets of user sequences whose top-10 lists generated by the LLM-based and CF-based models, respectively, contain the target item.

\newpage

When the intersection ratio is minimized ($\omega \approx 0$), $\lambda$ converges to 0.5, reflecting the orthogonality between the two models, and ensuring a balanced influence from both models. This adjustment improves the quality of the recommendation by effectively merging diverse perspectives from the LLM and CF-based method.
The analysis related to this approach is explained in Appendix \ref{apd:dpr}.

\section{Experiment}
This section presents the experimental results and analysis that demonstrate the effectiveness of our proposed method.

\begin{table*}[h]
    \centering
    \renewcommand{\arraystretch}{0.9}
    \resizebox{0.95\textwidth}{!}{
    \begin{tabular}{l|cccc|cccc|cccc}
        \toprule
        \multirow{2}{*}{Model} & \multicolumn{4}{c}{Sports} & \multicolumn{4}{c}{Beauty} & \multicolumn{4}{c}{Toys} \\
        \cmidrule(lr){2-5} \cmidrule(lr){6-9} \cmidrule(lr){10-13}
            & H@5 & N@5 & H@10 & N@10 & H@5 & N@5 & H@10 & N@10 & H@5 & N@5 & H@10 & N@10 \\
        \midrule
        SASRec             & \underline{0.0328} & \underline{0.0179} & \underline{0.0499} & \underline{0.0234} & \underline{0.0561} & \underline{0.0323} & \underline{0.0872} & \underline{0.0423} & \underline{0.0628} & \underline{0.0354} & \underline{0.0915} & \underline{0.0447} \\
        + LLMRank ICL      & 0.0201 & 0.0099 & 0.0341 & 0.0144 & 0.0316 & 0.0156 & 0.0555 & 0.0233 & 0.0410 & 0.0204 & 0.0636 & 0.0277 \\
        + LLMRank Seq      & 0.0187 & 0.0088 & 0.0328 & 0.0134 & 0.0279 & 0.0134 & 0.0506 & 0.0208 & 0.0388 & 0.0190 & 0.0624 & 0.0267 \\
        + LLMRank Recent   & 0.0198 & 0.0099 & 0.0335 & 0.0143 & 0.0304 & 0.0154 & 0.0525 & 0.0225 & 0.0415 & 0.0213 & 0.0639 & 0.0286 \\
        + TALLRec (w/o FT) & 0.0092 & 0.0046 & 0.0172 & 0.0072 & 0.0077 & 0.0038 & 0.0145 & 0.0060 & 0.0236 & 0.0122 & 0.0409 & 0.0178 \\
        + TALLRec (w FT)   & 0.0036 & 0.0028 & 0.0054 & 0.0033 & 0.0048 & 0.0029 & 0.0085 & 0.0041 & 0.0097 & 0.0069 & 0.0125 & 0.0078 \\
        + A-LLMRec         & 0.0200 & 0.0097 & 0.0435 & 0.0174 & 0.0063 & 0.0029 & 0.0311 & 0.0109 & 0.0493 & 0.0243 & 0.0911 & 0.0380 \\
        \midrule
        \textbf{\algname{} (+ SASRec)}  & \textbf{0.0367} & \textbf{0.0242} & \textbf{0.0571} & \textbf{0.0307} & \textbf{0.0675} & \textbf{0.0462} & \textbf{0.1022} & \textbf{0.0574} & \textbf{0.0856} & \textbf{0.0594} & \textbf{0.1189} & \textbf{0.0701} \\
        \midrule
        Improvement & 11.89\% & 35.20\% & 14.43\% & 31.20\% & 20.32\% & 43.03\% & 17.20\% & 35.70\% & 36.31\% & 67.80\% & 29.95\% & 56.82\% \\
        \midrule
        DIF-SR           & \underline{0.0360} & \underline{0.0197} & \underline{0.0542} & \underline{0.0256} & \underline{0.0576} & \underline{0.0336} & \underline{0.0901} & \underline{0.0442} & \underline{0.0700} & \underline{0.0404} & \underline{0.0995} & \underline{0.0499} \\
        + LLMRank ICL    & 0.0210 & 0.0104 & 0.0365 & 0.0154 & 0.0323 & 0.0159 & 0.0545 & 0.0231 & 0.0393 & 0.0197 & 0.0643 & 0.0278 \\
        + LLMRank Seq    & 0.0185 & 0.0088 & 0.0341 & 0.0138 & 0.0289 & 0.0139 & 0.0509 & 0.0210 & 0.0384 & 0.0189 & 0.0626 & 0.0268 \\
        + LLMRank Recent & 0.0204 & 0.0102 & 0.0356 & 0.0151 & 0.0318 & 0.0161 & 0.0532 & 0.0230 & 0.0433 & 0.0223 & 0.0656 & 0.0296 \\
        + TALLRec (w/o FT) & 0.0168 & 0.0095 & 0.0325 & 0.0145 & 0.0239 & 0.0166 & 0.0422 & 0.0224 & 0.0307 & 0.0173 & 0.0577 & 0.0260 \\
        + TALLRec (w FT)   & 0.0182 & 0.0150 & 0.0188 & 0.0152 & 0.0182 & 0.0150 & 0.0188 & 0.0152 & 0.0191 & 0.0115 & 0.0316 & 0.0156 \\
        + A-LLMRec         & 0.0240 & 0.0122 & 0.0449 & 0.0191 & 0.0093 & 0.0046 & 0.0328 & 0.0121 & 0.0362 & 0.0151 & 0.0896 & 0.0328 \\
        \midrule
        \textbf{\algname{} (+ DIF-SR)}  & \textbf{0.0371} & \textbf{0.0247} & \textbf{0.0590} & \textbf{0.0318} & \textbf{0.0699} & \textbf{0.0472} & \textbf{0.1032} & \textbf{0.0579} & \textbf{0.0885} & \textbf{0.0615} & \textbf{0.1228} & \textbf{0.0726}\\
        \midrule
        Improvement & 3.06\% & 25.38\% & 8.86\% & 24.22\% & 21.35\% & 40.48\% & 14.54\% & 31.00\% & 26.43\% & 52.23\% & 23.42\% & 45.49\%\\
        \midrule
    ELMRec             & \underline{0.0492} & \underline{0.0414} & \underline{0.0569} & \underline{0.0437} & \underline{0.0610} & \underline{0.0503} & \underline{0.0729} & \underline{0.0540} & \underline{0.0706} & \underline{0.0616} & \underline{0.0749} & \underline{0.0623} \\
        + LLMRank ICL      & 0.0229 & 0.0128 & 0.0312 & 0.0155 & 0.0290 & 0.0157 & 0.0409 & 0.0196 & 0.0296 & 0.0169 & 0.0367 & 0.0192 \\
        + LLMRank Seq      & 0.0154 & 0.0084 & 0.0279 & 0.0124 & 0.0236 & 0.0128 & 0.0355 & 0.0166 & 0.0279 & 0.0162 & 0.0366 & 0.0190 \\
        + LLMRank Recent   & 0.0167 & 0.0091 & 0.0272 & 0.0125 & 0.0236 & 0.0129 & 0.0361 & 0.0169 & 0.0275 & 0.0160 & 0.0346 & 0.0183 \\
        + TALLRec (w/o FT) & 0.0214 & 0.0122 & 0.0277 & 0.0142 & 0.0221 & 0.0128 & 0.0309 & 0.0156 & 0.0290 & 0.0173 & 0.0352 & 0.0194 \\
        + TALLRec (w FT)   & 0.0151 & 0.0086 & 0.0207 & 0.0104 & 0.0084 & 0.0047 & 0.0112 & 0.0056 & 0.0259 & 0.0155 & 0.0315 & 0.0173 \\
        + A-LLMRec         & 0.0007 & 0.0003 & 0.0026 & 0.0009 & 0.0000 & 0.0000 & 0.0004 & 0.0001 & 0.0005 & 0.0002 & 0.0012 & 0.0004 \\
        \midrule
        \textbf{\algname{} (+ ELMRec)}       & \textbf{0.0589} & \textbf{0.0455} & \textbf{0.0746} & \textbf{0.0506} & \textbf{0.0816} & \textbf{0.0625} & \textbf{0.1060} & \textbf{0.0704} & \textbf{0.1092} & \textbf{0.0777} & \textbf{0.1350} & \textbf{0.0860}\\
        \midrule
        Improvement & 19.72\% & 9.90\% & 31.11\% & 15.79\% & 33.77\% & 24.25\% & 45.40\% & 30.37\% & 54.67\% & 26.14\% & 80.24\% & 38.04\%\\
        \bottomrule
    \end{tabular}}
    \vspace{-1mm}
    \caption{Evaluation results in the LLM-recommender cooperation scenario. The best and second-best results for each metric are highlighted in bold and underlined, respectively. "H@k" and "N@k" denote Hit Rate and NDCG at rank k, respectively.}
    \label{tab:evaluation_results}
    \vspace{-5mm}
\end{table*}

\subsection{Experimental Setup}

To evaluate the effectiveness of \algname{}, we integrate existing CF-based recommendation method into our method and compared its performance against existing LLM-based reranking methods. 
Our goal was to examine whether the proposed parallel architecture would be more effective than the traditional serial reranking approach. 
For scalability, we employed LLaMA3.2 3B~\cite{llama} for query generation and the implementation of LLM-based baselines, while for similarity-based retrieval, we leveraged a widely used pre-trained encoder~\cite{encoder} without any additional training.

\fix{\ding{202} In the first experiment, we evaluated the performance of \algname{} in comparison to existing LLM-based reranking methods when adapting sequential recommendation models. 
For the existing LLM-based reranking methods, recommendation models are used as candidates selectors similar to the previous study}~\cite{palr}. 
\ding{203} In the second experiment, we compared \algname{} against existing CF-based approaches, T5-based approaches, and retrieval-based approaches that involve fine-tuning LLMs for query generation. 

To better reflect real-world recommendation scenarios with a large candidate pool, we evaluated performance over the entire item pool.
We adopted two widely used ranking metrics: \emph{Hit Rate (HR)} and \emph{Normalized Discounted Cumulative Gain (NDCG)}, with $k \in \{5, 10\}$.
More details on the experimental settings and implimentation are provided in the appendix and online repository\footnote{https://github.com/venzino-han/QueRec2025}.

\subsection{Main Results} 
In this subsection, we present key results comparing our method with existing baselines.

\subsubsection{LLM-recommender Cooperation Scenario}

\begin{table*}[h]
    \centering
    \renewcommand{\arraystretch}{0.9}
    \resizebox{0.98\textwidth}{!}{
    \begin{tabular}{l|l|cccc|cccc|cccc}
        \toprule
        \multirow{2}{*}{Method} & \multirow{2}{*}{Model} & \multicolumn{4}{c}{Sports} & \multicolumn{4}{c}{Beauty} & \multicolumn{4}{c}{Toys} \\
        \cmidrule(lr){3-6} \cmidrule(lr){7-10} \cmidrule(lr){11-14}
        & & H@5 & N@5 & H@10 & N@10 & H@5 & N@5 & H@10 & N@10 & H@5 & N@5 & H@10 & N@10 \\
        \midrule
         \multirow{8}{*}{CF-based} & Caser & 0.0116 & 0.0072 & 0.0194 & 0.0097 & 0.0205 & 0.0131 & 0.0347 & 0.0176 & 0.0166 & 0.0107 & 0.0270 & 0.0141 \\
          & GRU4Rec & 0.0129 & 0.0086 & 0.0204 & 0.0110 & 0.0164 & 0.0099 & 0.0283 & 0.0137 & 0.0097 & 0.0059 & 0.0176 & 0.0084 \\
          & HGN & 0.0189 & 0.0120 & 0.0313 & 0.0159 & 0.0325 & 0.0206 & 0.0512 & 0.0266 & 0.0321 & 0.0221 & 0.0497 & 0.0277 \\
          & SASRec & 0.0328 & 0.0179 & 0.0499 & 0.0234 & 0.0561 & 0.0323 & 0.0872 & 0.0423 & 0.0628 & 0.0354 & 0.0915 & 0.0447 \\
          & BERT4Rec & 0.0115 & 0.0075 & 0.0191 & 0.0099 & 0.0203 & 0.0124 & 0.0347 & 0.0170 & 0.0116 & 0.0071 & 0.0203 & 0.0099 \\
          & FDSA & 0.0182 &0.0122 &0.0288 &0.0156 &0.0267 &0.0163 &0.0407 &0.0208 &0.0228 &0.0140&0.0381 &0.0189 \\
          & S$^3$-Rec & 0.0251&0.0161&0.0385 &0.0204&0.0387 &0.0244 &0.0647&0.0327 &0.0443 &0.0294&0.0700 &0.0376 \\
          & DIF-SR  & 0.0360& 0.0197 & 0.0542 & 0.0256 &0.0576 & 0.0336 & 0.0901 &0.0442 & 0.0700 & 0.0404 &0.0995 & 0.0499 \\
          \midrule
          \multirow{5}{*}{T5-based}& P5      &  0.0272 & 0.0169 & 0.0361 & 0.0198 & 0.0503 & 0.0370 & 0.0659 & 0.0421 & 0.0648 & 0.0567 & 0.0709 & 0.0587 \\
          & TIGER  & 0.0264&  0.0181  & 0.0400 & 0.0225 &0.0454 & 0.0321 & 0.0648 &0.0384 & 0.0521 & 0.0371 &0.0712 & 0.0432 \\
          & POD      &  0.0496 & 0.0396 &  0.0576 &  0.0419 &  0.0537 &  0.0395 & 0.0688 & 0.0443 & 0.0691 & 0.0599 &  0.0742 &  0.0610 \\
          & RDRec & 0.0505 & 0.0408 & 0.0596 & 0.0433 & 0.0601 & 0.0461 & 0.0743 & 0.0504 & 0.0723 & 0.0593 & 0.0802 & 0.0605\\
          & ELMRec & 0.0492 & 0.0414 & 0.0569 & 0.0437 & 0.0610 & 0.0503 & 0.0729 & 0.0540 & 0.0706 & 0.0616 & 0.0749 & 0.0623\\
        \midrule
        \multirow{2}{*}{Retrieval-based} & GPT4Rec           & 0.0002&  0.0001  & 0.0004 & 0.0002 &0.0004 & 0.0002 & 0.0006 &0.0003 & 0.0005 & 0.0003 &0.0011 & 0.0005 \\
        & BIGRec  & 0.0071 & 0.0046 & 0.0115 & 0.0061 & 0.0237 & 0.0169 & 0.0355 & 0.0207 & 0.0324 & 0.0229 & 0.0471 & 0.0276 \\
        \midrule
        Ours & \textbf{\algname{}} & \textbf{0.0589} & \textbf{0.0455} & \textbf{0.0746} & \textbf{0.0506} & \textbf{0.0816} & \textbf{0.0625} & \textbf{0.1060} & \textbf{0.0704} & \textbf{0.1092} & \textbf{0.0777} & \textbf{0.1350} & \textbf{0.0860}\\
        \bottomrule
    \end{tabular}}
    \vspace{-1mm}
    \caption{Performance comparison of existing recommendation methods. The best results for each metric are highlighted in bold. "H@k" and "N@k" denote Hit Rate and NDCG at rank k, respectively.
    }
    \vspace{-1mm}

    \label{tab:exp2_results}
\end{table*}

The first experiment evaluated the effectiveness of our proposed parallel architecture. 
The results of this experiment are presented in Table~\ref{tab:evaluation_results}.
We observed that existing LLM-based reranking methods lead to a decrease in performance. 
This decline seems to be due to the fact that the selected candidate items from the CF-based model are very similar to each other, which may confuse the LLM during reranking. 
Although A-LLMRec, which uses CF-based embeddings, outperformed other LLM-based reranking methods, it still underperformed compared to our method.
These results suggest that learning CF-signals through natural language-based training remains a challenging task, even when the LLM is trained on candidate items from CF-based models.

In contrast, \algname{} achieved an average performance improvement of \perfavg{} and a maximum improvement of \perfmax{}. 
In particular, \algname{} does not require additional training when integrated into an existing recommendation system, highlighting its epandability and ease of deployment. 
These results demonstrate that \algname{} can effectively leverage the knowledge inherent in LLM, thereby providing differentiated recommendations from CF-based models.

\subsubsection{Compare to Traditional Recommenders}

In the second experiment, we compared the performance of \algname{} against traditional sequential recommendation models and LLM-based retrieval methods. 
The results presented in Table~\ref{tab:exp2_results} demonstrate that \algname{} outperforms existing recommendation models. 
Furthermore, \algname{} significantly outperformed existing LLM fine-tuning methods such as GPT4Rec and BIGRec, which generate item-titles and use them as queries. 
These results suggest that approaches based on fine-tuning may introduce biases induced by datasets and potentially compromise the inherent generalization capabilities of LLMs.

\begin{table*}[h]
    \renewcommand{\arraystretch}{0.9}
    \centering
    \resizebox{0.95\textwidth}{!}{
    \begin{tabular}{l|cccc|cccc|cccc}
        \toprule
        \multirow{2}{*}{Method} & \multicolumn{4}{c}{Sports} & \multicolumn{4}{c}{Beauty} & \multicolumn{4}{c}{Toys} \\
        \cmidrule(lr){2-5} \cmidrule(lr){6-9} \cmidrule(lr){10-13}
        & H@5 & N@5 & H@10 & N@10 & H@5 & N@5 & H@10 & N@10 & H@5 & N@5 & H@10 & N@10 \\
        \midrule
        \algname{} & \textbf{0.0589} & \textbf{0.0455} & \textbf{0.0746} & \textbf{0.0506} & \textbf{0.0816} & \textbf{0.0625} & \textbf{0.1060} & \textbf{0.0704} & \textbf{0.1092} & \textbf{0.0777} & \textbf{0.1350} & \textbf{0.0860}\\
        w/o CF-based model   & 0.0280 & 0.0188 & 0.0403 & 0.0228 & 0.0501 & 0.0354 & 0.0675 & 0.0410 & 0.0680 & 0.0472 & 0.0937 & 0.0555 \\
        w/o recent item info & 0.0274 & 0.0185 & 0.0407 & 0.0228 & 0.0459 & 0.0319 & 0.0622 & 0.0372 & 0.0625 & 0.0434 & 0.0908 & 0.0525 \\
        w/o item description & 0.0248 & 0.0165 & 0.0369 & 0.0204 & 0.0454 & 0.0317 & 0.0625 & 0.0373 & 0.0625 & 0.0423 & 0.0888 & 0.0508 \\
        w/o user query       & 0.0095 & 0.0061 & 0.0166 & 0.0084 & 0.0211 & 0.0131 & 0.0325 & 0.0168 & 0.0254 & 0.0162 & 0.0412 & 0.0213 \\
          
        \bottomrule
    \end{tabular}}
    \vspace{-2mm}
    \caption{Evaluation results of ablation study. The best results for each metric are highlighted in bold.
    }
    \label{tab:ablation}
    \vspace{-2mm}
\end{table*}

\subsection{In-depth Analysis}
In this subsection, we analyze the distinguishing advantages of the proposed method. 

\begin{figure*}[ht!]
    \centering


    \begin{minipage}[t]{0.32\textwidth}
        \centering
        {\includegraphics[width=\textwidth]{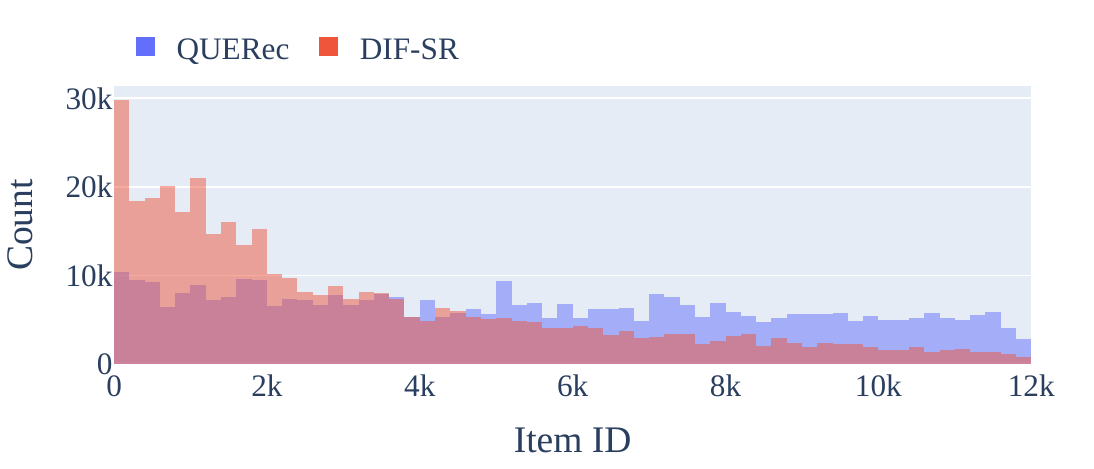}}
    \end{minipage}
    \begin{minipage}[t]{0.32\textwidth}
        \centering
        {\includegraphics[width=\textwidth]{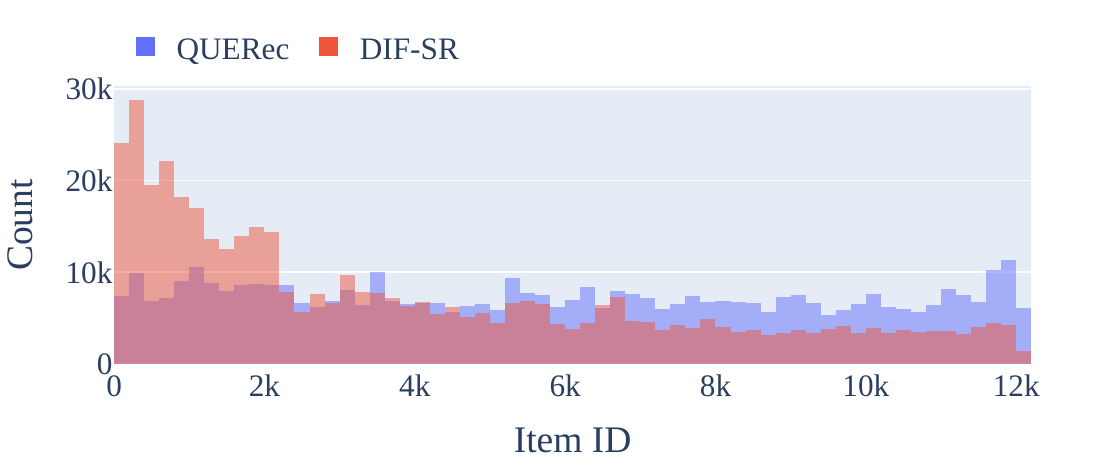}}
    \end{minipage}
    \begin{minipage}[t]{0.32\textwidth}
        \centering
        {\includegraphics[width=\textwidth]{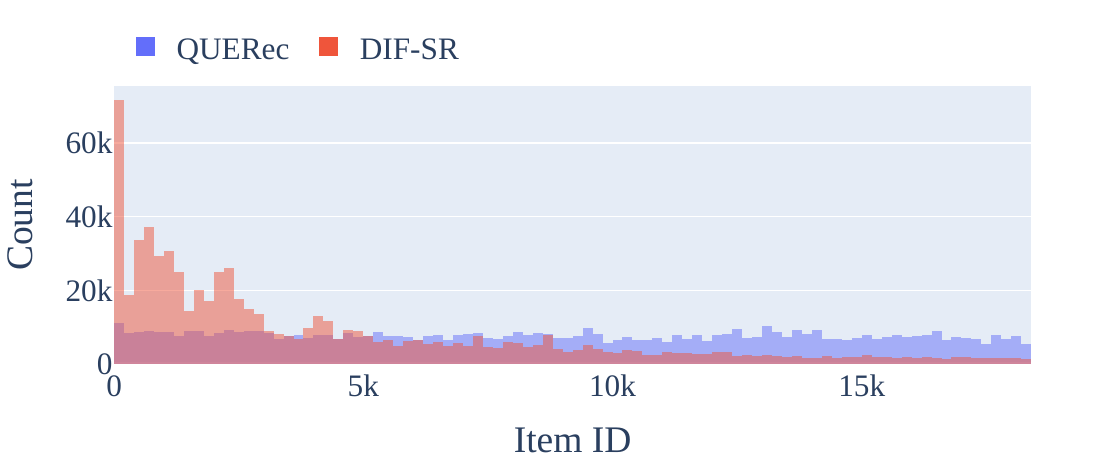}}
    \end{minipage}
    
    \vspace{-0.25cm}

    \begin{minipage}[t]{0.32\textwidth}
        \centering
        {\includegraphics[width=\textwidth]{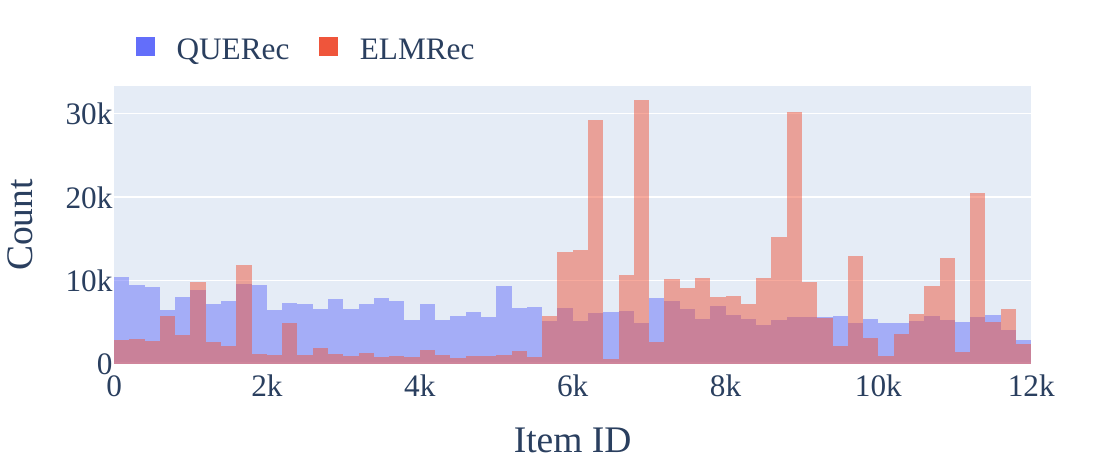}}
        \vspace{-0.7cm}
        \caption*{{\small (a) Toys.}}
    \end{minipage}
    \begin{minipage}[t]{0.32\textwidth}
        \centering
        {\includegraphics[width=\textwidth]{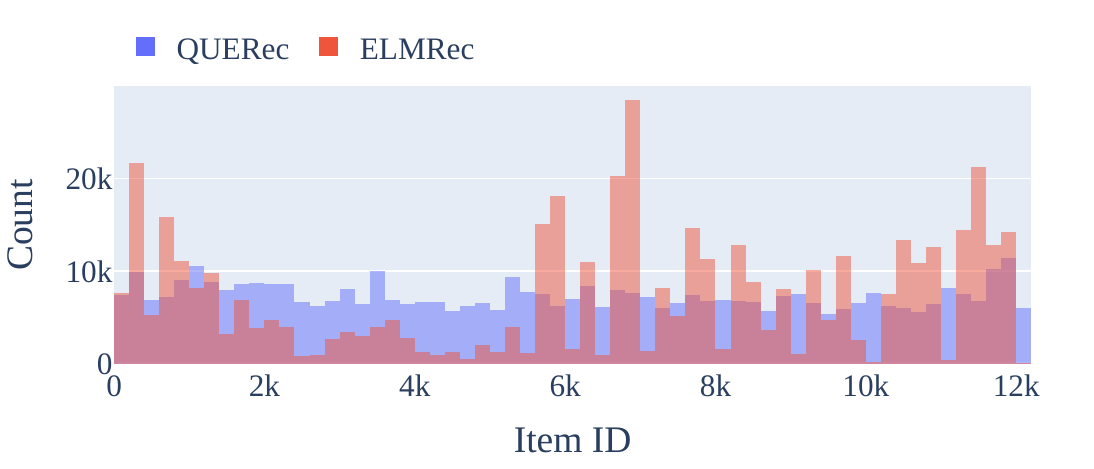}}
        \vspace{-0.7cm}
        \caption*{{\small (b) Beauty.}}
    \end{minipage}
    \begin{minipage}[t]{0.32\textwidth}
        \centering
        {\includegraphics[width=\textwidth]{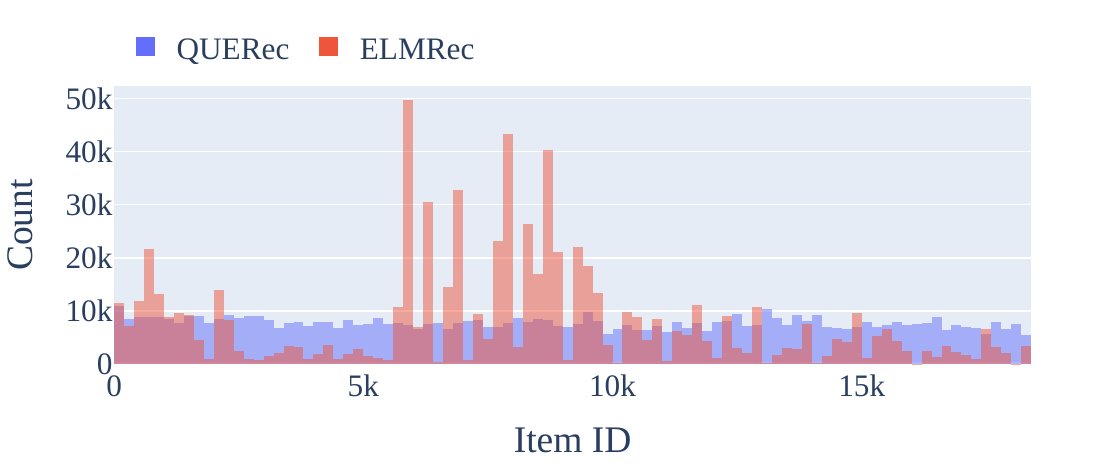}}
        \vspace{-0.7cm}
        \caption*{{\small (c) Sports.}}
    \end{minipage}
    
    \vspace*{-0.2cm}
    \caption{Diversity of recommendations across different datasets and models. The x-axis represents the IDs of the recommended items, while the y-axis indicates the frequency of recommendation. The top-20 recommended items for each user were extracted and compared.}
    \vspace*{-0.5cm}
    \label{fig:distribution_comparison}
\end{figure*}

\subsubsection{Ablation Study}

We conducted an ablation study to assess the contribution of each component in our approach. 
As shown in Table~\ref{tab:ablation}, we evaluated performance changes when each of the key components was removed: (1) CF-based model (2) recent item information, (3) generated item description, and (4) user queries. 
The absence of user queries resulted in the most significant performance degradation, while excluding the other components also led to notable declines. 
Another notable finding is that our method outperforms fine-tuned LLM-based approaches (GPT4Rec and BIGRec), even without being combined with a CF-based model. 
This result is encouraging, as our method does not require any LLM fine-tuning steps. 

\subsubsection{Diversity of Recommendations}
Our proposed method does not rely on additional training, allowing unbiased and more  diverse recommendations compared to traditional methods that often reinforce popularity bias through training. 
As shown in Fig.\ref{fig:distribution_comparison}, existing methods exhibit skewed item distributions, suggesting that the training process introduces biases into these models.
In contrast, our proposed method, produces a more balanced distribution of recommendation, indicating improved diversity.
%

This characteristic can mitigates filter bubbles and promotes exploratory user experiences. 
These findings highlight that our method not only incorporates a distinct perspective compared to traditional recommendation models, but also has the potential to generate synergies when integrated with existing approaches.
Further analysis is provided in Appendix\ref{apd_exp}.

\begin{figure*}[!th]
    \centering
    \includegraphics[width=0.97\textwidth]{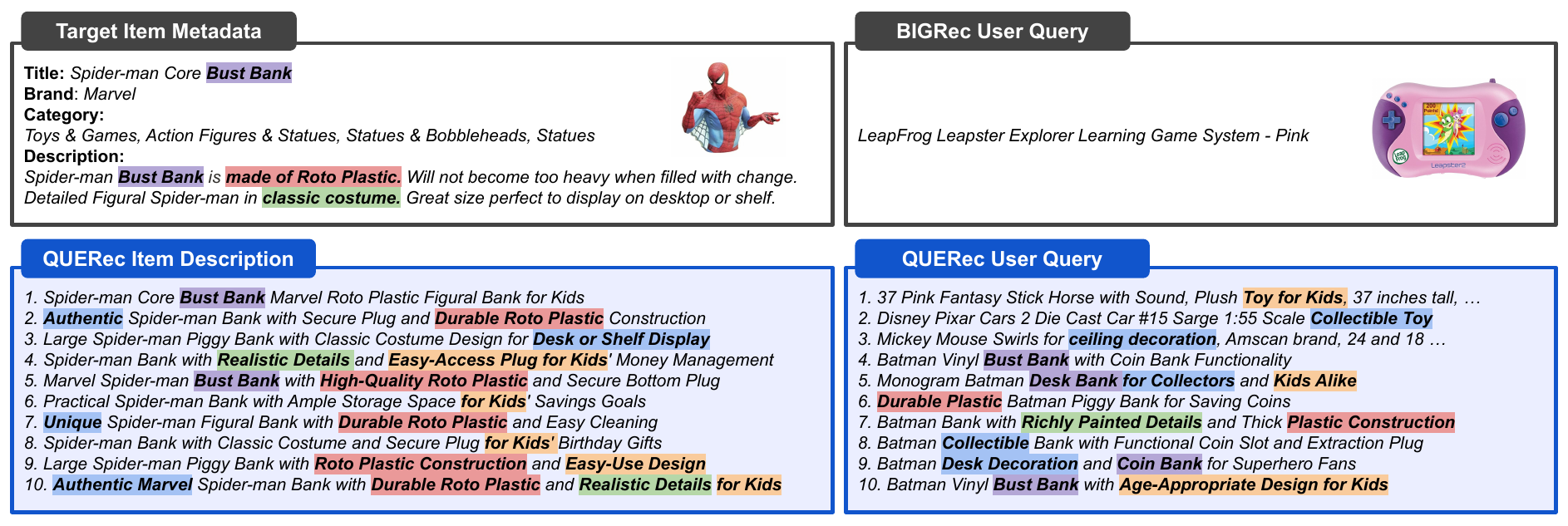}
    \vspace{-3mm}
    \caption{\fix{Query quality comparison in the Toys Dataset. Related attributes are highlighted in the same color.}}
    \label{fig:query_quality}
    \vspace{-3mm}
\end{figure*}

\begin{figure}[!h]
    \vspace{-1mm}
    \centering
    \includegraphics[width=0.95\linewidth]{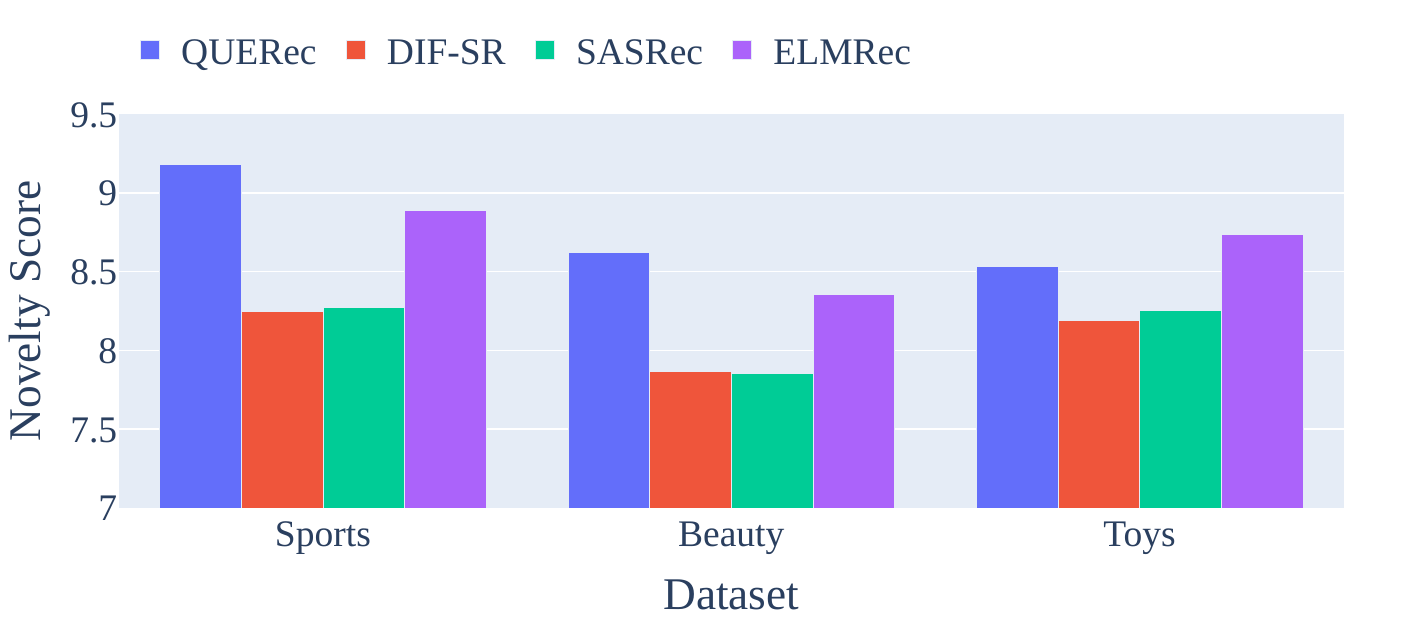}
    \vspace{-0.1cm}
    \caption{\fix{Novelty evaluation in three datasets.}}
    \label{fig:novelty_eval}
    \vspace{-0.1cm}
\end{figure}

\subsubsection{\fix{Novelty Evaluation}}

To examine whether the observed improvement in the diversity of recommendations translates into the recommendation of more novel items, we analyze based on the average novelty scores of items in the Hit@10 results. 
Novelty is computed as the negative logarithm of the relative item popularity~\cite{novelty}, defined as:
\begin{equation}
\text{Novelty}(i) = -\log \left( \frac{\text{freq}(i)}{\sum_{j \in \mathcal{I}} \text{freq}(j)} \right),
\end{equation}
\noindent where $\text{freq}(i)$ denotes the frequency of item $i$ in the training set, and $\mathcal{I}$ is the set of all items.

Fig.~\ref{fig:novelty_eval} presents the average novelty scores of the top-10 recommended items for each method. In most cases, our proposed method outperformed existing approaches, indicating its effectiveness in promoting the recommendation of less popular and more novel items.
These results suggest that our proposed method is capable of capturing fundamentally different signals compared to traditional recommendation systems, indicating its potential to deliver personalized recommendations tailored to users with diverse preferences.



\subsubsection{\fix{Query Quality Analysis}}

To examine whether the proposed method effectively leverages the broad knowledge encoded in LLMs, we qualitatively compare the generated queries. Existing approaches, which are fine-tuned to produce item-title-style queries, often exhibit dataset-induced biases and struggle to provide clear reasoning for recommendations. In contrast, our method incorporates user history and reviews into the query, enabling retrieval based on richer contextual information and offering interpretability in the form of implicit rationales.

Fig.~\ref{fig:query_quality} presents example queries generated by BIGRec and \algname{}, along with the actual items purchased by users. While \algname{} does not explicitly mention the exact item titles, its queries align with the purchased items in terms of target user groups and product characteristics. 
Notably, our method generates semantically relevant queries that reflect the preferences of the user. 
These findings indicate that LLMs can leverage their knowledge for accurate and unbiased recommendations without requiring fine-tuning.
Additional examples are included in Appendix \ref{apd_query_quality}.

\section{Conclusion}
In this work, we introduced \algname{}, a novel parallel framework that enhances the diversity and performance of recommendation by using LLM to generate personalized queries. 
\algname{} outperforms state-of-the-art baselines including LLM-based reranking, LLM-based next-item prediction, and CF-based methods, by effectively leveraging the rich knowledge embedded in LLMs.
Furthermore, our analysis highlights the ability of the method to reduce bias and increase the diversity of recommendations. 
By integrating the complementary strengths of LLMs and recommendation models, \algname{} provides a promising direction for future recommendation research, offering a practical, training-free solution adaptable to evolving LLMs and large-scale dynamic item pools.

\newpage

\section{Limitations}
Despite its advantages, \algname{} has several limitations. 
First, while it eliminates the need for explicit candidate selection, the effectiveness of query generation depends on the quality of the pre-trained LLM and its ability to generalize across domains. 
In cases where the LLM lacks sufficient domain-specific knowledge, the generated queries may not capture user intent effectively. 
Second, since our approach relies on retrieving items based on semantic similarity, it may underperform in domains where structured interaction data (e.g., collaborative signals) are more informative than textual item representations. 
Finally, while \algname{} enhances diversity, it does not explicitly control for fairness, popularity bias, or serendipity, which remain open challenges for future research. 
Addressing these limitations through hybrid approaches, enhanced query generation mechanisms, and domain-aware enhancements presents a promising direction for further improving LLM-driven recommendation systems.

\section{Ethics Statement}

\paragraph{Potential Risks}
Our study was conducted on fixed datasets and the potential impact of the user in real-world applications has not been examined. Therefore, caution is required when applying the proposed method beyond the controlled experimental setting.

\paragraph{Use of Scientific Artifacts}
Our research leveraged open source tools, including PyTorch \cite{pytorch}, along with pre-trained language models such as LLaMA3.2 and T5 obtained via the Huggingface \cite{huggingface} library. 
We used all artifacts in accordance with their intended purpose.

\paragraph{Use of Ai Assistants}
We only used ChatGPT to provide a better expression and to refine the language.
Some of the code used in the experiment was written with the assistance of Copilot.

\section{Acknowledgments}
This research was supported by the National Research Foundation of Korea (RS-2025-16071337).
For GPU infrastructure, our work was supported by the IITP grant funded by MSIT (No. RS-2025-02653113, High-Performance Research AI Computing Infrastructure Support at the 2 PFLOPS Scale).


\bibliography{cite}

\appendix
\onecolumn

\section{\fix{Clarification of contribution}}
In this section, we further elaborate on the aspects that distinguish our work from prior studies and clarify the key contributions of this paper.

Early research on LLM-based recommendation systems primarily relied on fine-tuning LLMs as recommendation models~\cite{recformer,hllm,tiger,p5}. These methods either trained LLMs to generate user/item embeddings or directly predict item IDs. However, they often underutilize LLMs’ reasoning abilities and incur substantial training costs due to full-parameter updates.
Another recent studies have proposed injecting LLM-derived user and item information into recommendation models to leverage LLM knowledge without fine-tuning the LLMs~\cite{kar,ur4rec}. 
However, these methods require retraining the recommendation model to align with the LLM outputs.
In contrast, we propose a training-free framework that fully leverages the pretrained LLM’s reasoning capacity. This design eliminates the need for fine-tuning, reducing bias and improving generalization.

Prior work such as GPT4Rec and BIGRec~\cite{gpt4rec,bigrec} fine-tuned LLMs to generate the title-based queries. This restricts the expressiveness of the LLM and limits adaptability in dynamic item pools. Our method, by contrast, generates rich, personalized queries that incorporate user preferences and item attributes without training, enhancing retrieval quality and recommendation diversity.

Recent reranking approaches~\cite{llmrank,tallrec,allmrec,palr} apply LLMs to reorder a small, pre-selected candidate set. However, these candidates are typically derived from CF-based models like SASRec~\cite{palr}, constraining the LLM’s capacity to explore diverse knowledge. We address this limitation by using LLMs to generate both user and item queries for full-pool retrieval, enabling knowledge-rich, unbiased, and adaptive recommendations.
Our proposed method also avoids the positional bias caused by the “lost in the middle” phenomenon, as it does not include candidate items in the prompt.

Another line of work involves using LLMs to generate synthetic queries, which are then used to train retrievers~\cite{mint,privacy_query}. These studies have primarily been evaluated on search tasks where user intent is explicitly expressed through natural language queries. In contrast, our work focuses on recommendation settings, where user intent is implicit and not directly observable. Unlike prior efforts that use synthetic queries to train retrievers, we aim to leverage the knowledge encoded in LLMs to enrich limited user-item information and apply it to retrieval-based recommendation. Furthermore, we propose a parallel architecture that effectively integrates this augmented information with existing recommender models without requiring retraining.



\section{Experimental Setups}
This section provides a more detailed description of the experimental setup and the baselines used.

\subsection{\fix{Environment}}
For the query generation stage using the LLM, we employed the vLLM\footnote{https://docs.vllm.ai/} framework, which enables efficient and scalable inference by leveraging optimized memory management and parallelization strategies.


\subsection{Datasets}

\begin{table}[!h]
    \centering
    
    \resizebox{0.5\columnwidth}{!}{
    \begin{tabular}{l|cccccc}
        \toprule
        \toprule
        Dataset & \#Users & \#Items & \#Reviews & Density (\%) \\
        \midrule
        Sports & 35,598 & 18,357 & 296,337 & 0.0453 \\
        Beauty & 22,363 & 12,101 & 198,502 & 0.0734 \\
        Toys & 19,412 & 11,924 & 167,597 & 0.0724   \\
        Yelp & 30,431 & 20,033 & 316,354 & 0.0519 \\
        \bottomrule
        \bottomrule
    \end{tabular}
    }
    \caption{Statistics of the datasets.\label{tab:dataset_statistics}}
\end{table}

We conducted experiments on four widely used benchmark datasets collected from the Amazon e-commerce platform: \emph{Sports \& Outdoors}, \emph{Beauty}, \emph{Toys \& Games} and \emph{Yelp}. 
Each dataset consists of user interactions, including a user ID, an item ID, a rating, a review, and a timestamp. 
The statistics of the data set are provided in Table~\ref{tab:dataset_statistics}. 
These datasets are widely adopted by previous studies~\cite{pod,elmrec,rdrec,tiger,aligning_knowledge,p5}, which have been extensively explored over the past three years. 
Due to space limitations, we include only the experimental results for the three main datasets in the manuscript, excluding the Yelp dataset.
To further validate the effectiveness of our proposed method, we conducted additional experiments on the Yelp dataset, which represents a domain distinct from the other three datasets. 
The results of experiment on the Yelp dataset are presented in Section~\ref{yelp}.

\subsection{Baselines \label{apd_baseline}}
This subsection introduces the baselines used in our experiments and their corresponding setups.
To obtain similarity scores for SASRec and DIF-SR, we trained the models using Recbole\footnote{https://recbole.io/}, a well-validated open-source recommendation framework. 
Detailed hyperparameters and training results will be made available through our public repository. 
For ELMRec, we excluded test targets affected by label leakage in the explanation task to ensure fair evaluation. In case of ELMRec, we obtain beam search scores and use them as the recommendation score.
The model was trained using the same T5-small backbone and hyperparameters as in prior studies. For all other baselines, we report the performance metrics as documented in previous studies \cite{s3rec,pod,elmrec}.
In cases of fine-tuned LLM-based methods (e.g. TALLRec, BIGRec and GPT4Rec), we trained the LLaMA 3.2 3B with LoRA for 2 epochs with a batch size of 4 and a learning rate of 5e-5.

\subsubsection{CF-base Model}
\begin{itemize}
    \item \textbf{Caser}~\cite{caser}: A CNN-based sequential recommendation model that learns item embeddings through convolution operations to effectively model users' sequential patterns.    
    \item \textbf{GRU4Rec}~\cite{gru4rec}: An RNN-based sequential recommendation model that leverages GRU to capture users' sequential patterns.
    \item \textbf{HGN}~\cite{hgn}: A sequential recommendation model that captures both long-term and short-term user interests through a hierarchical gating mechanism, which selectively processes item features and instances while explicitly modeling item-item relationships.
    \item \textbf{SASRec}~\cite{sasrec}: A sequential recommendation model that employs a self-attention mechanism to learn users' sequential behavior patterns.
    \item \textbf{BERT4Rec}~\cite{bert4rec}: A sequential recommendation model that utilizes BERT architecture to capture users' sequential interaction patterns.
    \item \textbf{FDSA}~\cite{fdsa}: A sequential recommendation model that captures both item-level and feature-level transition patterns by applying separate self-attention blocks to model their relationships, enhancing recommendation performance. It integrates heterogeneous item features using attention mechanisms and combines item and feature transitions to improve next-item prediction.
    \item \textbf{S$^3$-Rec}~\cite{s3rec}: A self-attentive model that enhances sequential recommendation by leveraging self-supervised pre-training objectives to capture correlations among attributes, items, subsequences, and sequences using mutual information maximization. 
    \item \textbf{DIF-SR}~\cite{sasrecd}: A sequential recommendation model that enhances side information fusion by shifting it from the input to the attention layer, decoupling attention calculations to mitigate rank bottlenecks and improve gradient flexibility, thereby enhancing modeling capacity and boosting recommendation performance.
\end{itemize}

\subsubsection{T5-base Model}
\begin{itemize}
    \item \textbf{P5}~\cite{p5}: A T5-based text-to-text recommendation model, converting all recommendation-related data into natural language sequences and utilizing personalized prompts for various recommendation tasks.
    \item \textbf{TIRGER}~\cite{tiger}: A generative model-based recommendation method, explicitly using numeric item-IDs to perform next-item prediction tasks. Unlike our approach, it does not leverage natural language generation, focusing instead on digit-based input-output structures.
    \item \textbf{POD}~\cite{pod}: A T5-based method, distills discrete prompts into continuous prompt vectors to better bridge user/item, aiming to improve both training efficiency and recommendation performance.
    \item \textbf{RDRec}~\cite{rdrec}: A T5-based method, enhances recommendation by distilling rationales from user and item reviews, allowing a compact model to leverage preference and attribute-level explanations for improved recommendation performance.
    \item \textbf{ELMRec}~\cite{elmrec}: A T5-based recommendation by enhancing whole-word embeddings to better interpret high-order user-item interactions without graph pre-training, while also addressing recency bias through a reranking mechanism to improve both direct and sequential recommendations.
\end{itemize}

\subsubsection{LLM-based Reranking Method}
\begin{itemize}
\item \textbf{LLMRank}~\cite{llmrank}: A zero-shot reranking approach using LLMs has been proposed to rerank selected candidate items. Three types of prompts were introduced: sequential, in-context learning, and recency-focused.
\item \textbf{TALLRec}~\cite{tallrec}: An approach that fine-tunes LLMs using LoRA to enhance their recommendation capabilities, enabling efficient adaptation to recommendation tasks while reducing computational overhead.
The initially proposed method used a binary prediction approach, whereas we modified the prompt to select the top-10 items from the candidate set used in a previous study~\cite{allmrec}. 
During training, the prompt included the top-20 candidate items recommended by the traditional recommendation model, while the training target was a reordered list of 10 item titles, ensuring that the label item appeared first. 
This setup was designed to allow a fair comparison that closely aligns with our proposed method.
\item \textbf{A-LLMRec}~\cite{allmrec}: An LLM-based framework that integrates pre-trained CF-based user/item embeddings into LLMs, enabling effective recommendations in both cold and warm scenarios.
It leverages both collaborative signals and LLM reasoning capabilities by introducing a alignment layer that requires training to bridge the two representations.

\end{itemize}

\subsubsection{LLM-based Retrieval Method}
\begin{itemize}
    \item \textbf{GPT4Rec}~\cite{gpt4rec}: An LLM-based recommendation method that fine-tunes language models to predict a user’s next item and uses the generated item text as search queries. 
    
    \item \textbf{BIGRec}~\cite{bigrec}: A bi-step grounding method for LLM-based recommendation systems. It first aligns the LLM with the recommendation task by training it to generate item titles. These generated titles are then used as textual queries to retrieve items, thereby grounding the language model outputs to the recommendation space.

\end{itemize}



\section{Experiments on Various LLMs}
To evaluate the generalizability of our proposed approach across different LLM architectures, we conducted experiments using various models, including LLaMA 3.1 8B, LLaMA 3.3 70B, Gemma 2 2B, and Gemma 2 3B. Table~\ref{tab:llm2} presents the performance of \algname{} in a standalone setting.
For the 70B model, we utilized \emph{Llama-3.3-70B-Instruct-Turbo} in DeepInfra\footnote{https://deepinfra.com/} for generation.

The results demonstrate that our method maintains stable performance across different LLMs, confirming its adaptability to various model architectures. Additionally, we observe that even lightweight models can achieve performance comparable to larger models, highlighting the efficiency and scalability of our approach.

\begin{table*}[h]
    \centering
    \resizebox{\textwidth}{!}{
    \begin{tabular}{l|cccc|cccc|cccc}
        \toprule
        \multirow{2}{*}{Model} & \multicolumn{4}{c}{Sports} & \multicolumn{4}{c}{Beauty} & \multicolumn{4}{c}{Toys} \\
        \cmidrule(lr){2-5} \cmidrule(lr){6-9} \cmidrule(lr){10-13}
        & H@5 & N@5 & H@10 & N@10 & H@5 & N@5 & H@10 & N@10 & H@5 & N@5 & H@10 & N@10 \\
        \midrule
        LLaMA 3.2 3B  & 0.0280 & 0.0188 & 0.0403 & 0.0228 & 0.0501 & 0.0354 & 0.0675 & 0.0410 & 0.0680 & 0.0472 & 0.0937 & 0.0555 \\
        LLaMA 3.1 8B  &  0.0270 & 0.0185 & 0.0397 & 0.0225 & 0.0491 & 0.0346 & 0.0667 & 0.0402 & 0.0665 & 0.0466 & 0.0926 & 0.0550 \\
        LLaMA 3.3 70B &  0.0269 & 0.0182 & 0.0399 & 0.0224  & 0.0505 & 0.0361 & 0.0686 & 0.0419 & 0.0678 & 0.0469 & 0.0935 & 0.0552\\
        Gemma 2 2B    &  0.0272 & 0.0186 & 0.0395 & 0.0225 & 0.0496 & 0.0356 & 0.0682 & 0.0416 & 0.0661 & 0.0457 & 0.0927 & 0.0543 \\        
        Gemma 2 9B    &  0.0247 & 0.0165 & 0.0390 & 0.0211 & 0.0481 & 0.0345 & 0.0680 & 0.0409 & 0.0664 & 0.0460 & 0.0940 & 0.0549 \\
        \bottomrule
    \end{tabular}}
    \caption{Performance variation of \algname{} across different LLMs. "H@k" and "N@k" denote Hit Rate and NDCG at rank k, respectively.}
    \label{tab:llm2}
\end{table*}


To investigate whether previously proposed reranking methods become more effective when utilizing a larger model, we conducted experiments using the LLaMA 3.3 70B. Table~\ref{tab:70B} presents the results of these experiments. Compared to lightweight models, we observed an overall improvement in performance compare to the cases used LLaMA 3.2 3B, and in some metrics, performance improvements were also observed compared to the standalone recommendation model.
However, in most cases, performance degradation was still evident, reaffirming that the reranking approach remains less effective than our proposed method. These findings highlight that simply scaling up the model does not necessarily resolve the limitations of existing reranking strategies.

\begin{table*}[h]
    \centering
    \resizebox{\textwidth}{!}{
    \begin{tabular}{l|cccc|cccc|cccc}
        \toprule
        \multirow{2}{*}{Model} & \multicolumn{4}{c}{Sports} & \multicolumn{4}{c}{Beauty} & \multicolumn{4}{c}{Toys} \\
        \cmidrule(lr){2-5} \cmidrule(lr){6-9} \cmidrule(lr){10-13}
            & H@5 & N@5 & H@10 & N@10 & H@5 & N@5 & H@10 & N@10 & H@5 & N@5 & H@10 & N@10 \\
        \midrule
        DIF-SR           & \underline{0.0360} & {0.0197} & \underline{0.0542} & {0.0256} & \underline{0.0576} & {0.0336} & \underline{0.0901} & \underline{0.0442} & {0.0700} & {0.0404} & \underline{0.0995} & {0.0499} \\
        + LLMRank ICL    & 0.0315 & 0.0215 & 0.0409 & 0.0246 & 0.0453 & 0.0322 & 0.0501 & 0.0339 & 0.0663 & 0.0478 & 0.0843 & 0.0537\\
        + LLMRank Seq    & 0.0348 & 0.0246 & 0.0431 & 0.0274 & 0.0489 & 0.0361 & 0.0534 & 0.0376 & 0.0725 & 0.0546 & 0.0891 & 0.0601\\
        + LLMRank Recent & 0.0354 & \underline{0.0252} & 0.0435 & \underline{0.0279} & 0.0491 & \underline{0.0364} & 0.0540 & 0.0380 & \underline{0.0739} & \underline{0.0556} & {0.0904} & \underline{0.0610}\\
        \midrule
        \algname{} (+ DIF-SR) & \textbf{0.0389} & \textbf{0.0243} & \textbf{0.0604} & \textbf{0.0312} & \textbf{0.0715} & \textbf{0.0479} & \textbf{0.1059} & \textbf{0.0589} & \textbf{0.0868} & \textbf{0.0618} & \textbf{0.1231} & \textbf{0.0735} \\
        \bottomrule
    \end{tabular}}
    \caption{Experimental results on LLaMA 3.3 70B. The best and second-best results for each metric are highlighted in bold and underlined, respectively. "H@k" and "N@k" denote Hit Rate and NDCG at rank k, respectively.}
    \label{tab:70B}
\end{table*}

\section{\fix{Divergent Perspective Reranking} \label{apd:dpr}}
In this section, we present the analysis and experiments that motivate our proposed Divergent Perspective Reranking approach. Through these studies, we demonstrate that our method offers greater stability and robustness compared to traditional ensemble techniques such as convex combination (CC) or reciprocal rank fusion (RRF).

\subsection{Intersection Between Recommendation Methods \label{apd:intersection}}

\begin{figure}[!h]
    \centering
    \vspace*{-0.5cm}
    \includegraphics[width=0.8\linewidth]{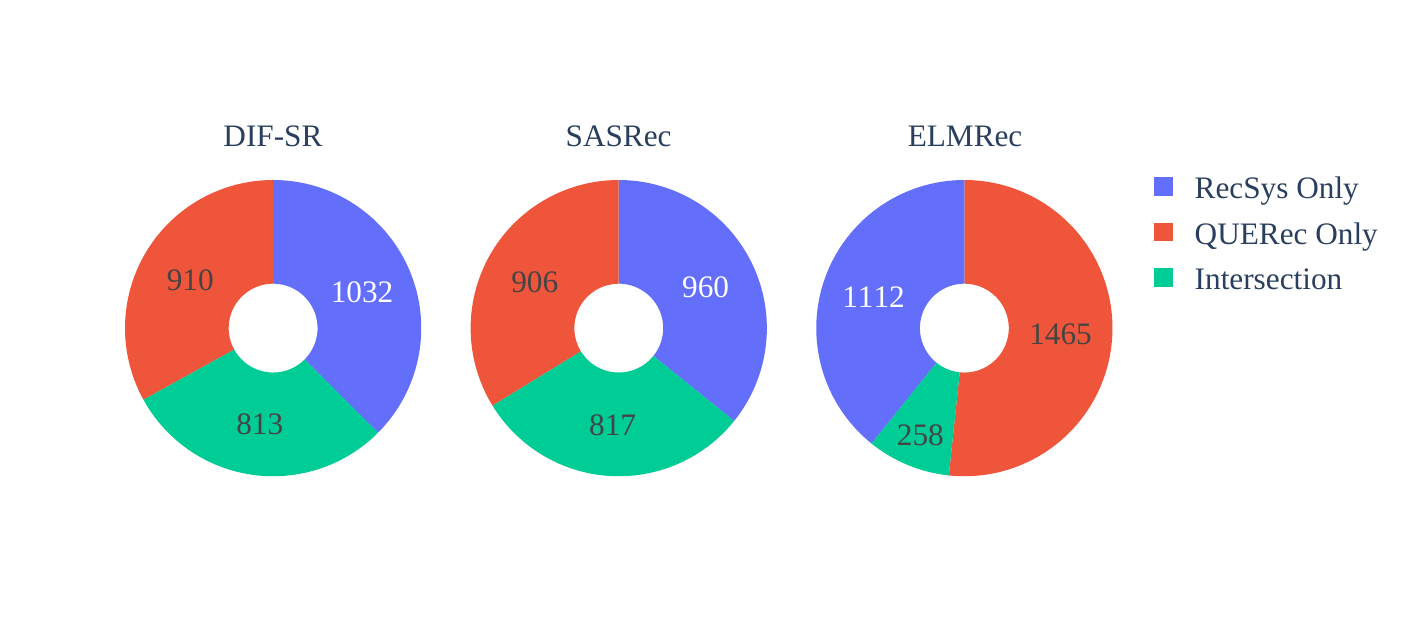}
    \vspace*{-1.2cm}
    \caption{Intersection between recommendation methods in Toys dataset.}
    \label{fig:inter_toys}
    \vspace*{-0.5cm}
\end{figure}
\begin{figure}[!h]
    \centering
    \includegraphics[width=0.8\linewidth]{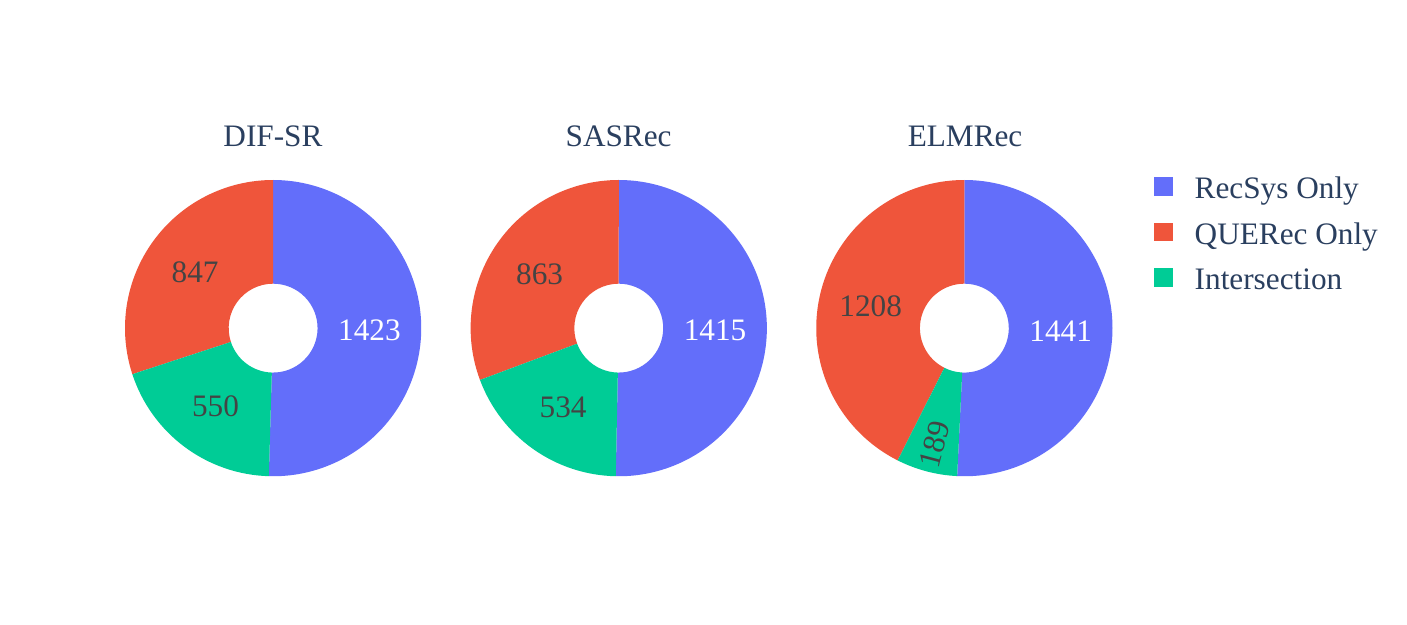}
    \vspace*{-1.2cm}
    \caption{Intersection between recommendation methods in Beauty dataset.}
    \label{fig:inter_beauty}
    \vspace*{-0.5cm}
\end{figure}
\begin{figure}[!h]
    \centering
    \includegraphics[width=0.8\linewidth]{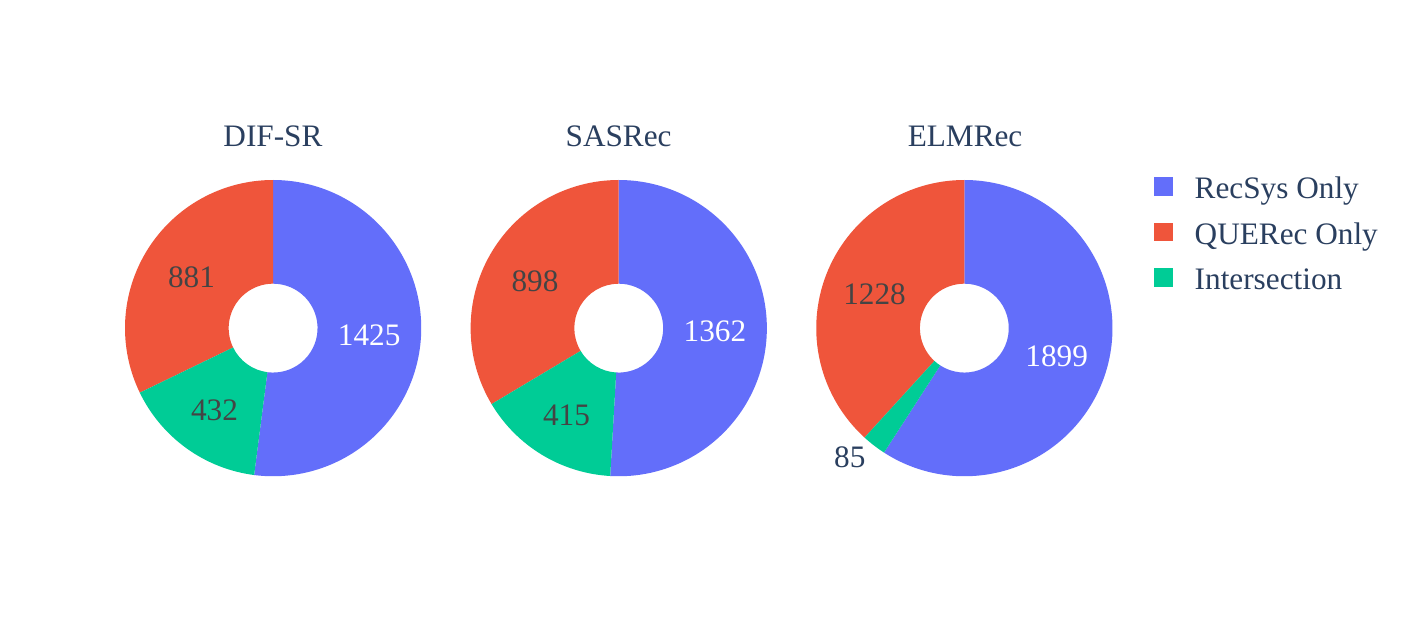}
    \vspace*{-1.2cm}
    \caption{Intersection between recommendation methods in Sports dataset.}
    \label{fig:inter_sports}
    \vspace*{-0.5cm}
\end{figure}

In this subsection, we present an analysis that motivated our proposed adaptive weighting scheme. Specifically, we examine the overlap in the recommended items between different recommendation methods, based on whether the recommended items are hits under the Hit@10 metric. Fig.\ref{fig:inter_toys}, \ref{fig:inter_beauty}, and \ref{fig:inter_sports} illustrate how the degree of intersection varies across methods and datasets. When the intersection is high, assigning weights based on the accuracy of each model can help improve the overall recommendation performance. However, when the intersection is low, skewed weights may lead to scenarios where incorrect recommendations from one model disproportionately influence the final result. To address this issue, we observe that using uniform weights (e.g., 0.5) in low-intersection cases can better optimize the final performance. Based on this insight, we propose an adaptive weighting strategy that considers not only the individual performance of each model but also the degree of intersection between their recommended items.

\begin{table*}[h]
    \centering
    \resizebox{\textwidth}{!}{
    \begin{tabular}{l|cccc|cccc|cccc}
        \toprule
        \multirow{2}{*}{Model} & \multicolumn{4}{c}{Sports} & \multicolumn{4}{c}{Beauty} & \multicolumn{4}{c}{Toys} \\
        \cmidrule(lr){2-5} \cmidrule(lr){6-9} \cmidrule(lr){10-13}
& H@5 & N@5 & H@10 & N@10 & H@5 & N@5 & H@10 & N@10 & H@5 & N@5 & H@10 & N@10 \\
\midrule
        
SASRec                & 0.0328 & 0.0179 & 0.0499 & 0.0234 & 0.0561 & 0.0323 & 0.0872 & 0.0423 & 0.0628 & 0.0354 & 0.0915 & 0.0447 \\
+ \algname{} CC (Ours)& \underline{0.0367} &    \textbf{0.0242} & \underline{0.0571} &    \textbf{0.0307} &    \textbf{0.0675} &    \textbf{0.0462} &    \textbf{0.1022} &    \textbf{0.0574} &   \textbf{0.0856} &    \textbf{0.0594}&   \textbf{0.1189} &   \textbf{0.0701} \\
+ \algname{} CC (0.9) &            0.0272  &            0.0179  &            0.0388  &            0.0216  &            0.0488  &            0.0340  &            0.0673  &            0.0400  &            0.0681 &            0.0458 &            0.0949 &            0.0545 \\
+ \algname{} CC (0.7) &            0.0317  &            0.0209  &            0.0452  &            0.0253  &            0.0565  &            0.0391  &            0.0816  &            0.0472  &            0.0796 &            0.0538 &            0.1089 &            0.0632 \\
+ \algname{} CC (0.5) &            0.0347  & \underline{0.0231} &            0.0534  & \underline{0.0291} &            0.0638  & \underline{0.0434} & \underline{0.0988} & \underline{0.0547} & \underline{0.0837}& \underline{0.0588}& \underline{0.1180}& \underline{0.0699} \\
+ \algname{} CC (0.3) &   \textbf{0.0371}  &            0.0219  &    \textbf{0.0576} &            0.0285  & \underline{0.0655} &            0.0421  & \underline{0.0988} &            0.0529  &            0.0774 &            0.0521 &            0.1107 &            0.0628 \\
+ \algname{} CC (0.1) &            0.0345  &            0.0192  &            0.0522  &            0.0249  &            0.0597  &            0.0360  &            0.0907  &            0.0460  &            0.0675 &            0.0401 &            0.0978 &            0.0498 \\
+ \algname{} RRF      &            0.0330  &            0.0198  &            0.0526  &            0.0261  &            0.0585  &            0.0374  &            0.0899  &            0.0475  &            0.0730 &            0.0473 &            0.1056 &            0.0578 \\
\midrule
DIF-SR & 0.0360 & 0.0197 & 0.0542 & 0.0256 & 0.0576 & 0.0336 & 0.0901 & 0.0442 & 0.0700 & 0.0404 & 0.0995 & 0.0499 \\
+ \algname{} CC (Ours)& \underline{0.0371}&    \textbf{0.0247} & \textbf{0.0590} & \textbf{0.0318} & \textbf{0.0699} & \textbf{0.0472} & \textbf{0.1032} & \textbf{0.0579} & \textbf{0.0885} & \textbf{0.0615} & \textbf{0.1228} & \textbf{0.0726} \\
+ \algname{} CC (0.9) &            0.0274 &            0.0180  & 0.0389 & 0.0217 & 0.0489 & 0.0340 & 0.0676 & 0.0401 & 0.0690 & 0.0461 & 0.0949 & 0.0545 \\
+ \algname{} CC (0.7) &            0.0327 &            0.0214  & 0.0465 & 0.0259 & 0.0580 & 0.0402 & 0.0833 & 0.0482 & 0.0804 & 0.0541 & 0.1098 & 0.0635 \\
+ \algname{} CC (0.5) &            0.0360 & \underline{0.0239} & 0.0557 & \underline{0.0303} & \underline{0.0668} & \underline{0.0457} & 0.0982 & \underline{0.0558} & \underline{0.0842} & \underline{0.0594} & \underline{0.1204} & \underline{0.0711} \\
+ \algname{} CC (0.3) &    \textbf{0.0380}&            0.0227  & \underline{0.0580} & 0.0292 & 0.0654 & 0.0437 & \underline{0.1003} & 0.0550 & 0.0770 & 0.0537 & 0.1120 & 0.0649 \\
+ \algname{} CC (0.1) &            0.0346 &            0.0194  & 0.0542 & 0.0257 & 0.0622 & 0.0381 & 0.0910 & 0.0473 & 0.0680 & 0.0427 & 0.0998 & 0.0530 \\
+ \algname{} RRF & 0.0338 & 0.0202 & 0.0532 & 0.0264 & 0.0608 & 0.0390 & 0.0920 & 0.0491 & 0.0731 & 0.0483 & 0.1073 & 0.0593 \\
\midrule
ELMRec & 0.0492 & 0.0414 & 0.0569 & 0.0437 & 0.0610 & 0.0503 & 0.0729 & 0.0540 & 0.0706 & 0.0616 & 0.0749 & 0.0623 \\
+ \algname{} CC (Ours) & \textbf{0.0589} & \underline{0.0455} & \textbf{0.0746} & \textbf{0.0506} & \textbf{0.0816} & \textbf{0.0625} & \textbf{0.1060} & \textbf{0.0704} & \textbf{0.1092} & \underline{0.0777} & \textbf{0.1350} & \underline{0.0860} \\
+ \algname{} CC (0.9) & 0.0256 & 0.0170 & 0.0381 & 0.0211 & 0.0477 & 0.0331 & 0.0648 & 0.0386 & 0.0658 & 0.0445 & 0.0919 & 0.0529 \\
+ \algname{} CC (0.7) & 0.0337 & 0.0217 & 0.0532 & 0.0279 & 0.0592 & 0.0401 & 0.0867 & 0.0489 & 0.0831 & 0.0564 & 0.1172 & 0.0674 \\
+ \algname{} CC (0.5) & \underline{0.0586} & \textbf{0.0462} & 0.0717 & \underline{0.0504} & \underline{0.0805} & \underline{0.0617} & \underline{0.1026} & \underline{0.0689} & \underline{0.1032} & \textbf{0.0811} & \underline{0.1267} & \textbf{0.0887} \\
+ \algname{} CC (0.3) & 0.0504 & 0.0427 & 0.0594 & 0.0456 & 0.0665 & 0.0545 & 0.0821 & 0.0595 & 0.0786 & 0.0676 & 0.0936 & 0.0724 \\
+ \algname{} CC (0.1) & 0.0491 & 0.0444 & 0.0562 & 0.0421 & 0.0623 & 0.0515 & 0.0745 & 0.0555 & 0.0681 & 0.0621 & 0.0721 & 0.0634 \\
+ RRF & 0.0576 & 0.0447 & \underline{0.0720} & 0.0494 & 0.0791 & 0.0595 & 0.1014 & 0.0667 & 0.0957 & 0.0752 & 0.1208 & 0.0833 \\

\bottomrule
\end{tabular}}
\caption{Experimental results based on various CF-LLM cooperation reranking methods. The best and second-best results for each metric are highlighted in bold and underlined, respectively. (Ours) denotes our proposed Hit Rate-based reranking method, while the other numerical values represent fixed lambda ($\lambda$) values used in the linear combination approach. "H@k" and "N@k" denote Hit Rate and NDCG at rank k, respectively.}
\label{tab:reranking}
\end{table*}

To integrate the top-k selection results from both the traditional recommendation model and the LLM-based approach, we employed a linear combination method. To prevent performance imbalance between the two models from negatively affecting the final ranking, we determined the $\lambda$ value based on validation Hit@10 scores and intersection ratios. To evaluate the effectiveness of this approach, we compared the performance of our method against two alternatives: fixing the $\lambda$ value and employing Reciprocal Rank Fusion (RRF). Table~\ref{tab:reranking} presents the results of these experiments.

The results indicate that our proposed method achieved the highest performance in most cases. 
However, in the case of ELMRec, the RRF-based approach demonstrated superior results. 
This suggests that text-to-text-based recommendation models generate distinct ranking patterns compared to traditional recommendation models, highlighting the need for further research on hybrid recommendation methods that dynamically incorporate multiple recommendation systems. Nonetheless, our proposed method remains a practical solution that can be stably applied across various models with minimal overhead.

\section{Additional Analysis}

\subsection{Item Distribution \label{apd_exp}}

\begin{figure*}[ht!]
    \centering
    \vspace{-0.4cm}
    \begin{minipage}[t]{0.32\textwidth}
        \centering
        {\includegraphics[width=\textwidth]{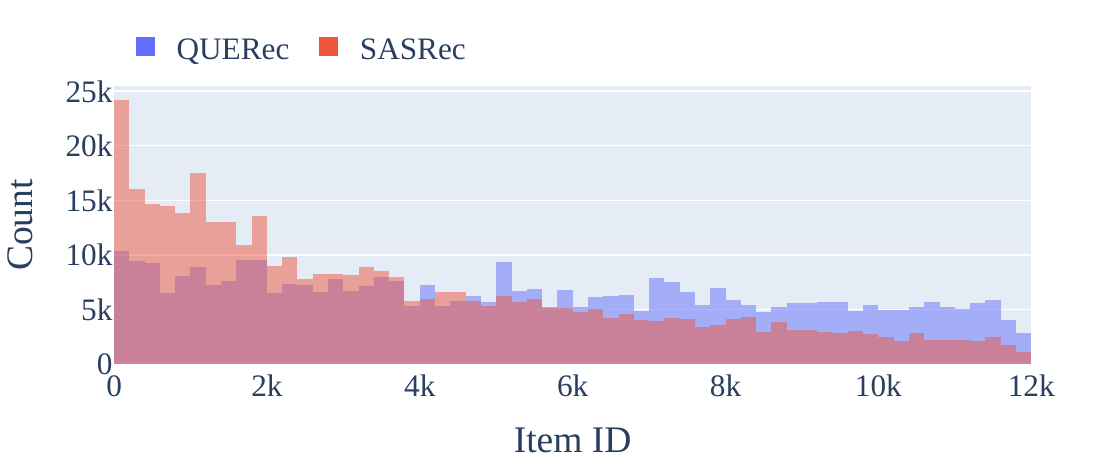}}
    \end{minipage}
    \begin{minipage}[t]{0.32\textwidth}
        \centering
        \centering
        {\includegraphics[width=\textwidth]{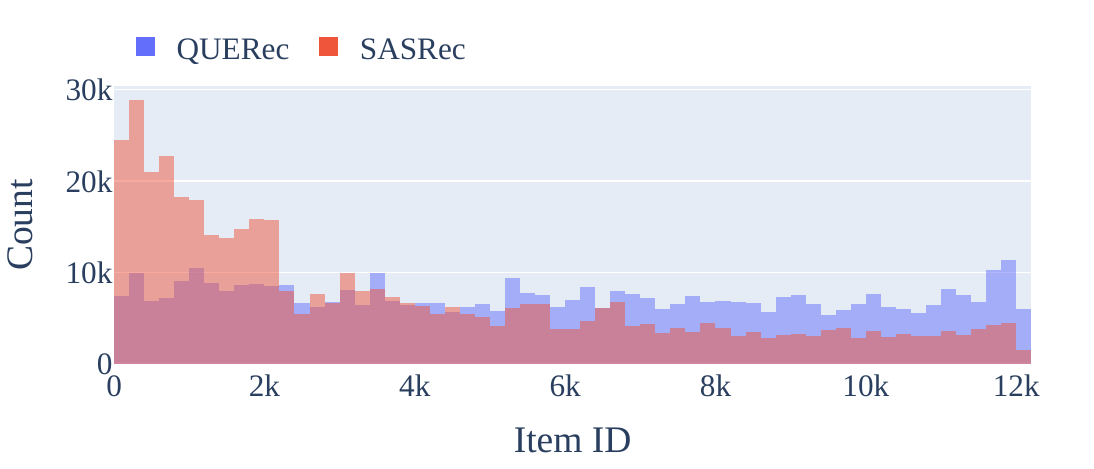}}
    \end{minipage}
    \begin{minipage}[t]{0.32\textwidth}
        \centering
        {\includegraphics[width=\textwidth]{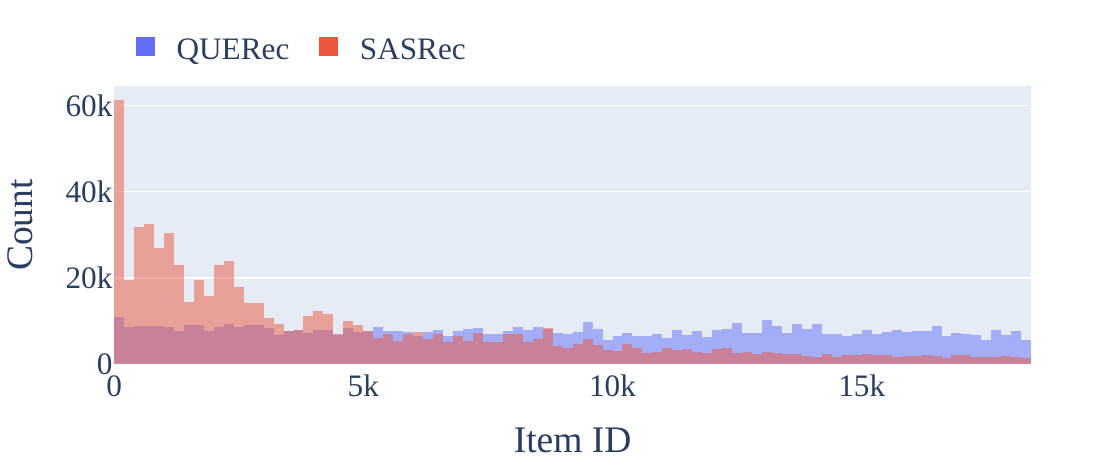}}
    \end{minipage}

    \vspace{-0.25cm}

    \begin{minipage}[t]{0.32\textwidth}
        \centering
        {\includegraphics[width=\textwidth]{fig/histogram_SASRecD_toys.pdf}}
    \end{minipage}
    \begin{minipage}[t]{0.32\textwidth}
        \centering
        {\includegraphics[width=\textwidth]{fig/histogram_SASRecD_beauty.pdf}}
    \end{minipage}
    \begin{minipage}[t]{0.32\textwidth}
        \centering
        {\includegraphics[width=\textwidth]{fig/histogram_SASRecD_sports.pdf}}
    \end{minipage}
    
    \vspace{-0.25cm}

    \begin{minipage}[t]{0.32\textwidth}
        \centering
        {\includegraphics[width=\textwidth]{fig/histogram_ELMRec_toys.pdf}}
        \vspace{-0.7cm}
        \caption*{{\small (a) Toys.}}
    \end{minipage}
    \begin{minipage}[t]{0.32\textwidth}
        \centering
        {\includegraphics[width=\textwidth]{fig/histogram_ELMRec_beauty.pdf}}
        \vspace{-0.7cm}
        \caption*{{\small (b) Beauty.}}
    \end{minipage}
    \begin{minipage}[t]{0.32\textwidth}
        \centering
        {\includegraphics[width=\textwidth]{fig/histogram_ELMRec_sports.pdf}}
        \vspace{-0.7cm}
        \caption*{{\small (c) Sports.}}
    \end{minipage}
    
    \caption{Diversity of recommendations across different datasets and models. The x-axis represents the IDs of the recommended items, while the y-axis indicates the frequency of recommendation. The top-20 recommended items for each user were extracted and compared.}
    \label{apd:distribution_comparison}
\end{figure*}

We stated that our proposed method has the potential to mitigate bias introduced by the training dataset. 
Fig.~\ref{apd:distribution_comparison} illustrates that major baselines exhibit strong bias toward certain items, whereas our method provides more balanced recommendations across a wide range of items. 
To investigate whether this bias originates from the training data, we conducted an additional analysis to examine the distribution of item interactions in the training set.

\begin{figure*}[t!]
    \centering
    \begin{minipage}[t]{0.32\textwidth}
        \centering
        {\includegraphics[width=\textwidth]{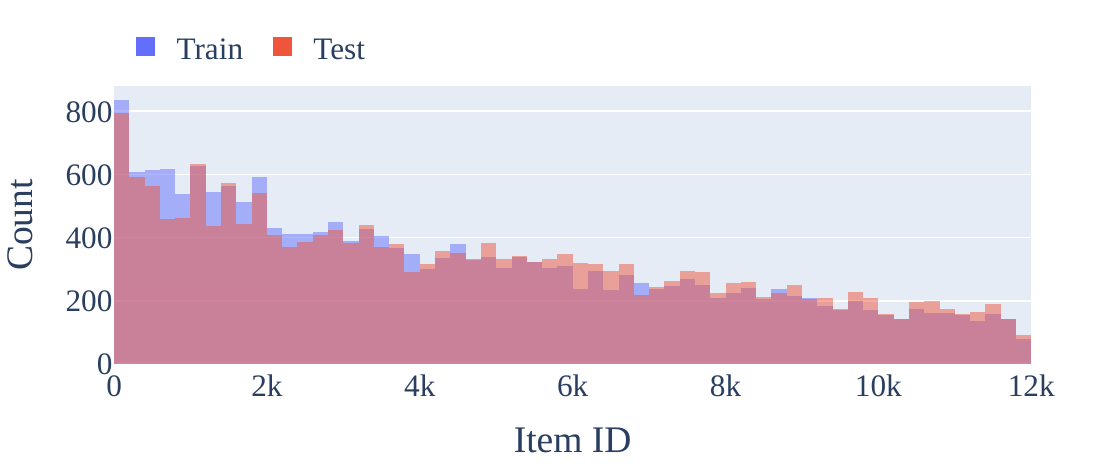}}
        \caption*{{\small (a) Toys.}}
    \end{minipage}
    \begin{minipage}[t]{0.32\textwidth}
        \centering
        \centering
        {\includegraphics[width=\textwidth]{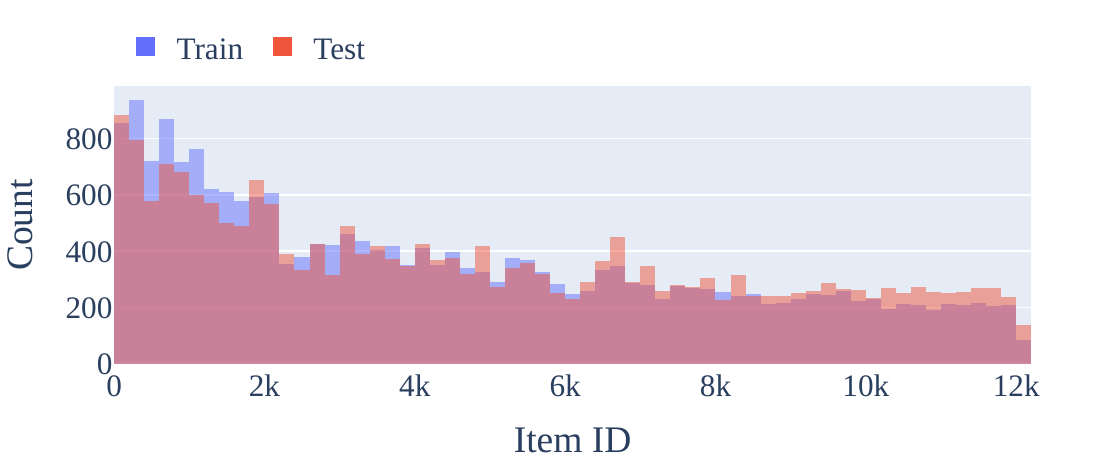}}
        \caption*{{\small (b) Beauty.}}
    \end{minipage}
    \begin{minipage}[t]{0.32\textwidth}
        \centering
        {\includegraphics[width=\textwidth]{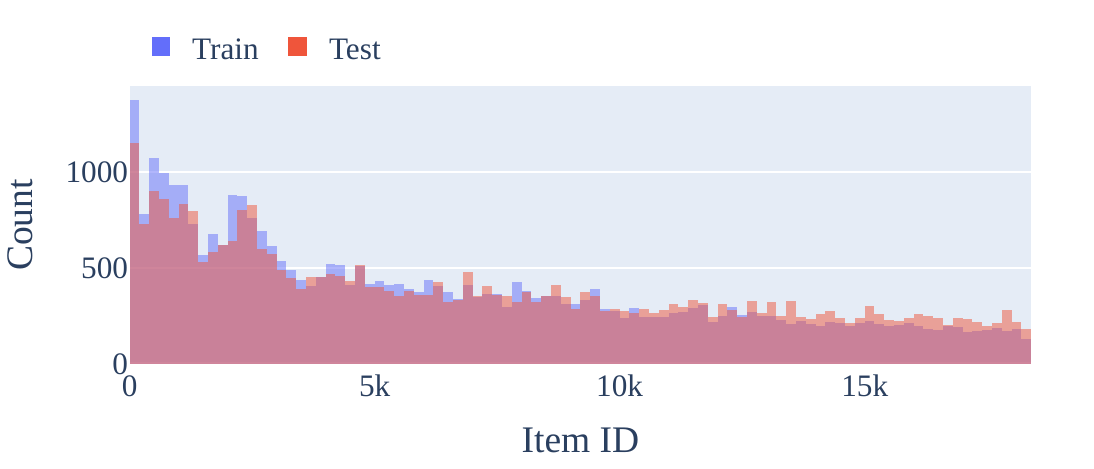}}
        \caption*{{\small (c) Sports.}}
    \end{minipage}

    \vspace{-0.25cm}
    \caption{Target item ID distribution across different datasets. The x-axis represents the IDs of the target items, while the y-axis indicates the frequency of items.}
    \vspace*{-0.3cm}
    \label{fig:item_dist}
\end{figure*}

We conducted a more detailed analysis of the bias present in conventional recommendation systems. Fig.~\ref{fig:item_dist} compares the distribution of item IDs in the training and test datasets used in our experiments. While the overall distributions are similar, we observe that the training dataset is more skewed towards lower item IDs, whereas the test dataset contains a relatively higher frequency of items with larger IDs.

\begin{figure}[!t]
    \centering
    \includegraphics[width=0.8\linewidth]{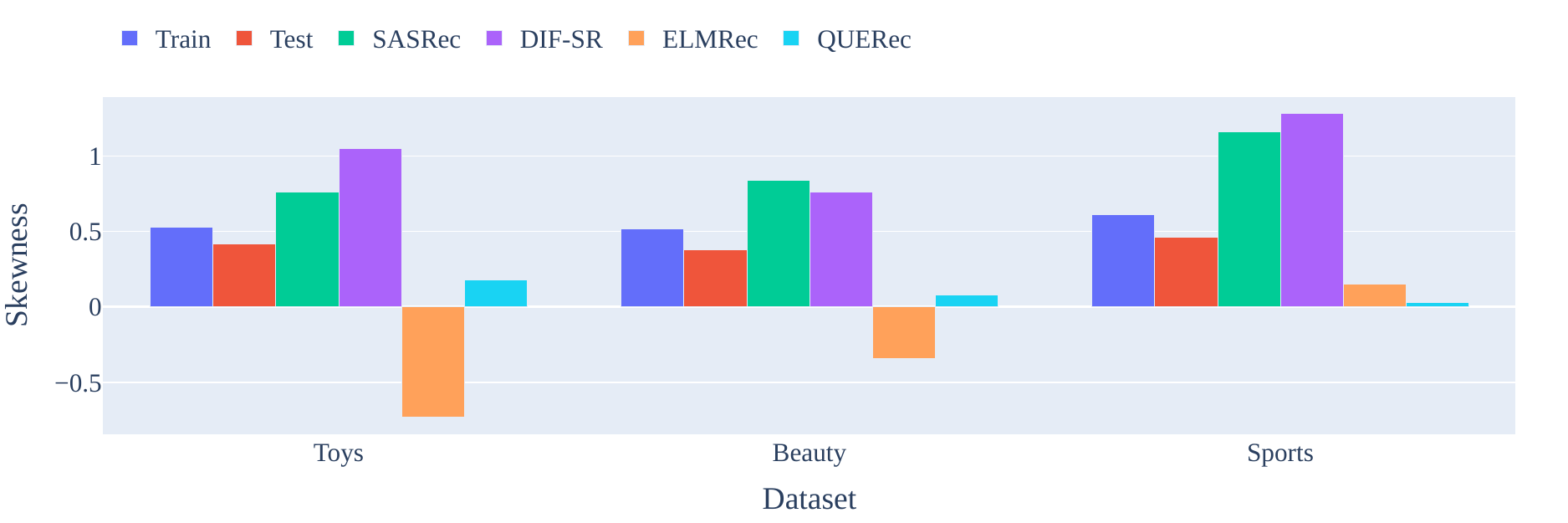}
    \caption{Comparison of item distribution skewness across datasets and the skewness of recommendation results for each model.}
    \label{fig:skewness}
    \vspace*{-0.2cm}
\end{figure}

To quantify this observation, we measured the \emph{skewness} of the distributions. Fig.~\ref{fig:skewness} presents a comparison of the skewness values for both datasets and the recommendation outputs of various models. Our analysis confirms that the training set exhibits a higher degree of skewness compared to the test set. Additionally, traditional recommendation models demonstrate an even greater level of skewness in their recommendation results than the training dataset itself. 

These findings indicate that conventional recommendation approaches may contribute to reduced diversity in recommendations. This further underscores the necessity of unbiased recommendation methods, such as our proposed approach, to mitigate such biases and improve recommendation fairness.

\subsection{\fix{Latency}}
We measured the latency of our system by generating recommendations for 100 users under our experimental setup. On average, query generation required 69.7 seconds, and dense retrieval using an encoder took 2.8 seconds, resulting in an overall latency of approximately 0.72 seconds per user. We consider this latency reasonable and acceptable for real-world recommendation systems.

While prior LLM-based methods may exhibit similar latency when handling a small candidate set, they become inefficient in realistic settings where the number of candidate items exceeds 10,000. These approaches typically include candidate items directly within the prompt, which significantly increases the input length and degrades performance during inference.
To mitigate this, such methods must partition the candidate pool into smaller subsets (e.g., fewer than 20 items per prompt) and perform multiple inference passes. However, this strategy introduces latency that grows linearly with the size of the candidate pool, rendering it unsuitable for deployment in large-scale recommender systems.

In contrast, our method decouples candidate selection from LLM inference, maintaining constant latency regardless of the size of the item pool. This key property makes our approach highly efficient and scalable, offering a practical solution for real-world recommendation scenarios.

\subsection{\fix{Query Quality Evaluation}\label{apd_query_quality}}
In this section, we evaluate the quality of the generated queries through both quantitative and qualitative analyses.

\begin{figure*}[!th]
    \centering
    \includegraphics[width=0.9\textwidth]{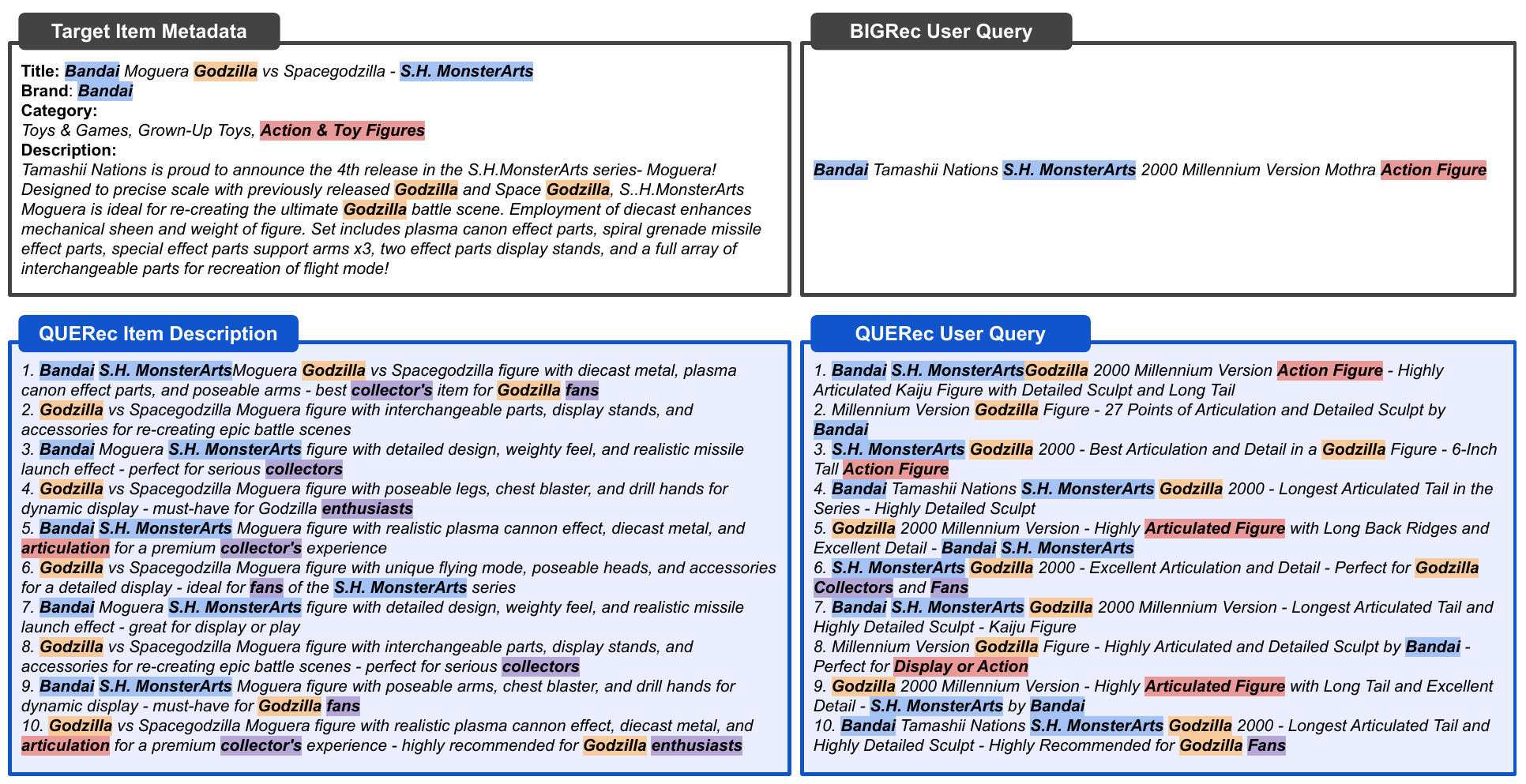}
    \vspace{-3mm}
    \caption{\fix{Query quality comparison in the Toys Dataset. Related attributes are highlighted in the same color.}}
    \label{fig:toy_case}
    \vspace{-5mm}
\end{figure*}
\begin{figure*}[!th]
    \centering
    \includegraphics[width=0.9\textwidth]{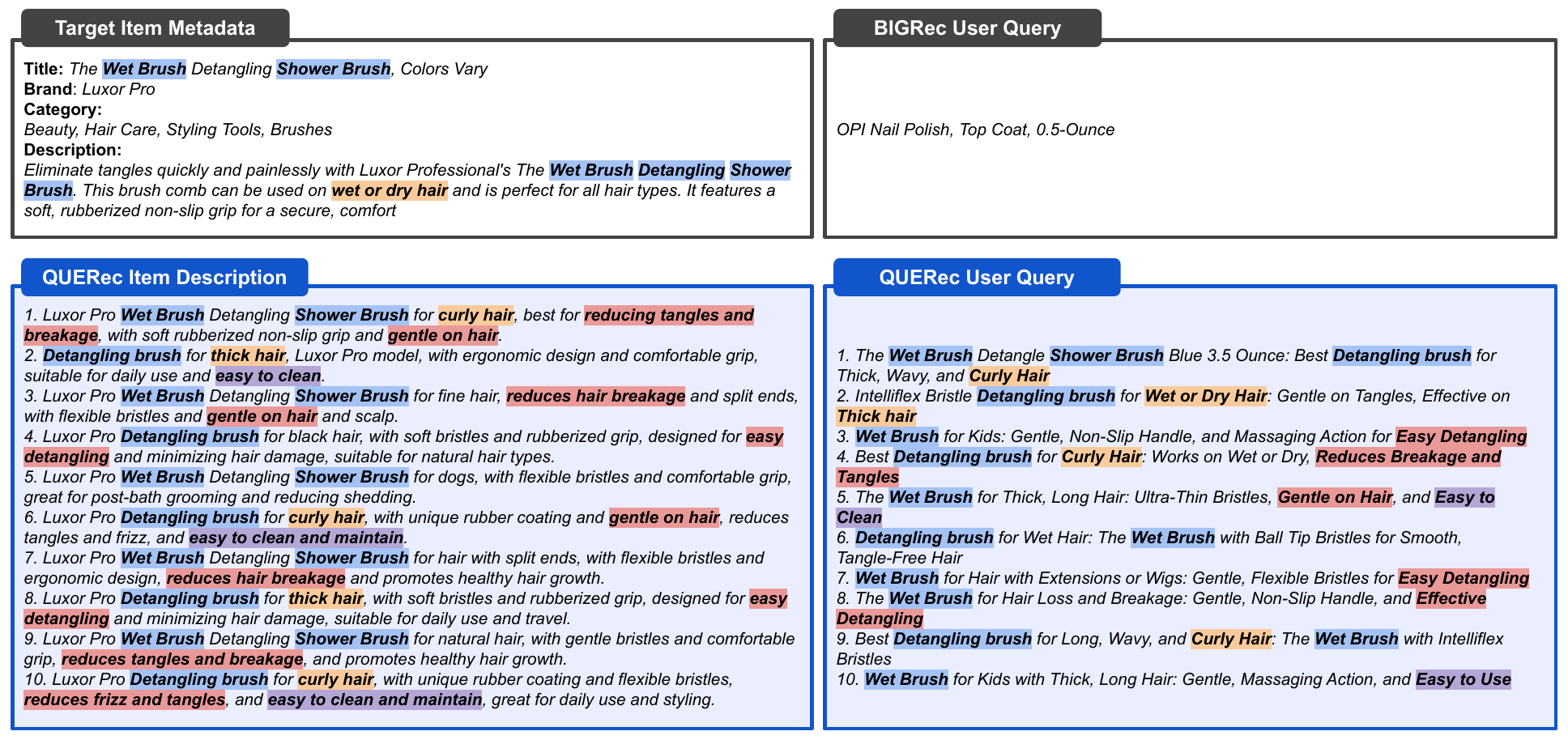}
    \vspace{-3mm}
    \caption{\fix{Query quality comparison in the Beauty Dataset. Related attributes are highlighted in the same color.}}
    \label{fig:beauty_case}
    \vspace{-5mm}
\end{figure*}
\begin{figure*}[!th]
    \centering
    \includegraphics[width=0.9\textwidth]{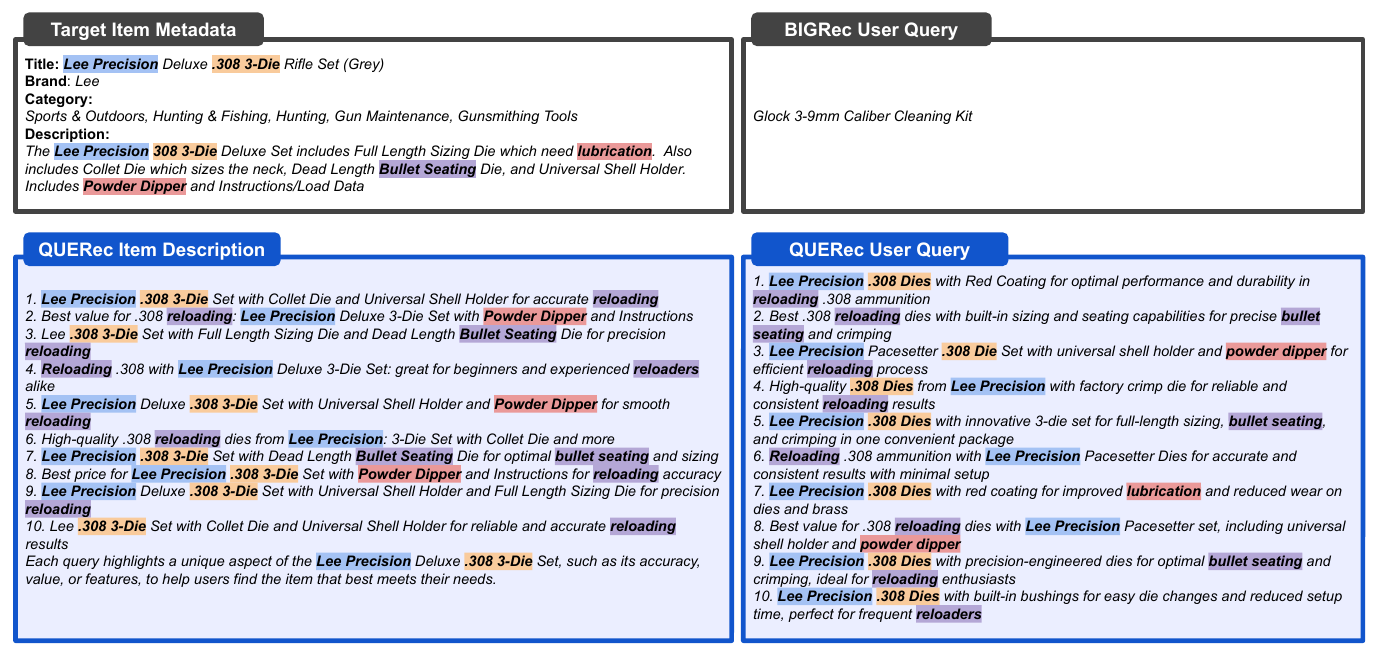}
    \vspace{-3mm}
    \caption{\fix{Query quality comparison in the Sports Dataset. Related attributes are highlighted in the same color.}}
    \label{fig:sports_case}
    \vspace{-5mm}
\end{figure*}

\subsubsection{\fix{Generated Query Comparision}\label{apd_query_quality}}

Fig. \ref{fig:toy_case}, \ref{fig:beauty_case}, and \ref{fig:sports_case} present examples from each dataset, showcasing the personalized user queries, expanded item descriptions, and corresponding target item information. Our proposed method successfully matches key product attributes with user preferences by generating richer information without any additional training. In contrast, fine-tuned baseline approaches often align with only limited features or mention irrelevant item titles. These results highlight the effectiveness of our method in leveraging the capabilities of LLMs for recommendation.

\subsubsection{\fix{Query Refinement}\label{apd_query_refine}}
To examine the potential impact of query quality on recommendation performance, we conducted additional experiments involving the removal of redundant or irrelevant queries. Specifically, we utilized LLaMA 3.3 70B to filter and refine the generated queries for 3,000 users. The evaluation results are summarized in Table~\ref{tab:query_refinement}.

\begin{table}[h]
    \centering
    \resizebox{\linewidth}{!}{
    \begin{tabular}{l|cccc|cccc|cccc}
        \toprule
        \multirow{2}{*}{Query Type} & \multicolumn{4}{c}{Sports} & \multicolumn{4}{c}{Beauty} & \multicolumn{4}{c}{Toys} \\
        \cmidrule(lr){2-5} \cmidrule(lr){6-9} \cmidrule(lr){10-13}
        & H@5 & N@5 & H@10 & N@10 & H@5 & N@5 & H@10 & N@10 & H@5 & N@5 & H@10 & N@10 \\
        \midrule
        Original & 0.0253 & 0.0173 & 0.0367 & 0.0209 & 0.0360 & 0.0252 & 0.0470 & 0.0287 & 0.0630 & 0.0446 & 0.0883 & 0.0528 \\
        Refined  & 0.0263 & 0.0173 & 0.0367 & 0.0205 & 0.0350 & 0.0247 & 0.0480 & 0.0289 & 0.0613 & 0.0434 & 0.0873 & 0.0518 \\
        \bottomrule
    \end{tabular}}
    \caption{Comparison of original and refined queries across datasets.}
    \label{tab:query_refinement}
\end{table}

Overall, the performance of refined queries were similar to the original ones, with minimal differences across most datasets. Notably, a slight performance drop was observed in the Toys domain. These findings suggest that the original queries produced by the LLM were already of high quality, providing effective signals for retrieval.

One possible explanation for the performance degradation is the loss of feature redundancy during the refinement process. In the original queries, certain attributes appeared multiple times, which may have implicitly reflected strong user preferences. By removing these repetitions, the refinement may have unintentionally weakened the representation of dominant user interests, thereby reducing the effectiveness of the final recommendations.

\section{\fix{Evaluation on Yelp Dataset}\label{yelp}}

We utilized datasets widely adopted by previous studies \cite{elmrec,tiger,pod,aligning_knowledge,rdrec,p5}, which have been extensively explored in recent research over the past three years. Major publications frequently employed the three datasets we selected. To ensure reproducibility and fair comparison, we chose the same datasets, believing they adequately demonstrate the effectiveness of our proposed approach.

To further validate the generalizability of our approach, we conducted additional experiments on the \emph{Yelp} dataset\footnote{https://www.yelp.com/dataset}. Unlike the previously used Amazon datasets, which are primarily focused on e-commerce domains (e.g., Beauty, Sports, Toys), Yelp consists of user-generated reviews centered around local businesses such as restaurants and services, presenting a distinct domain with different user behavior patterns and item characteristics. This expansion allows us to examine the adaptability of our method beyond typical product recommendation settings.
Yelp has been widely adopted in recent recommendation studies \cite{screc,s3rec,eager,idrec}, making it a reliable benchmark for evaluating model effectiveness across domains. 

Our method continues to demonstrate competitive or superior performance on this dataset, further supporting its applicability in diverse real-world scenarios. These results confirm that the advantages of our training-free, modular architecture—such as performance gains and improved diversity—are not confined to a specific dataset or domain, but extend to varied recommendation environments.

    

\begin{table}[h]
    \centering
    \resizebox{0.5\linewidth}{!}{
    \begin{tabular}{l|cccc}
        \toprule
        \multirow{2}{*}{Model} & \multicolumn{4}{c}{Yelp} \\
        \cmidrule(lr){2-5}
        & H@5 & N@5 & H@10 & N@10 \\
        \midrule
        Caser & 0.0151 & 0.0096 & 0.0253 & 0.0129 \\
        GRU4Rec & 0.0152 & 0.0099 & 0.0263 & 0.0134 \\
        HGN & 0.0186 & 0.0115 & 0.0326 & 0.0159 \\
        SASRec & 0.0452 & 0.0334 & 0.0630 & 0.0391 \\
        BERT4Rec & 0.0051 & 0.0033 & 0.0090 & 0.0045 \\
        FDSA & 0.0271 & 0.0170 & 0.0464 & 0.0232 \\
        S$^3$-Rec & 0.0168 & 0.0123 & 0.0341 & 0.0168 \\
        DIF-SR & 0.0452 & 0.0335 & 0.0651 & 0.0398 \\
        P5    & 0.0225 & 0.0159 & 0.0329 & 0.0193 \\
        TIGER & 0.0212 & 0.0146 & 0.0367 & 0.0194 \\
        \midrule
        QUEREC (+ SASRec)& \underline{0.0506} & \underline{0.0408} & \underline{0.0684} & \underline{0.0465} \\
        QUEREC (+ DIF-SR) & \textbf{0.0511} & \textbf{0.0410} & \textbf{0.0700} & \textbf{0.0471} \\
        QUEREC (w/o CF-based model) & 0.0314 & 0.0267 & 0.0376 & 0.0287 \\
        \bottomrule
    \end{tabular}}
    \caption{Performance comparison on the Yelp dataset. The best and second-best results for each metric are highlighted in bold and underlined, respectively. "H@k" and "N@k" denote Hit Rate and NDCG at rank k, respectively.}
    \label{tab:yelp_results}
\end{table}

\newpage

\section{Prompt and Generated Query}
\label{apd_prompt}
This section presents the prompts used in our query generation approach along with examples of the generated queries.
We designed the prompt format based on the LLaMA3 instruction format. For user history, we set the maximum length to 8 and included a phrase to emphasize the last interaction. To prevent excessive prompt length, we incorporated item queries only for the last interaction during user query generation. 
We instructed the LLMs to generate 10 queries per each item and user.
We utilized vLLM~\footnote{https://github.com/vllm-project/vllm} to perform the generation steps required for training-free methods, including query generation.
The complete set of generated queries will be released through an online public repository upon acceptance.

\subsection{Item Query Generation Prompt}

\begin{lstlisting}[language=json]
<|begin_of_text|><|start_header_id|>system<|end_header_id|>
You are an intelligent assistant designed to create detailed and precise search queries for items based on their descriptions and aggregated user reviews.
Your task is to generate 10 distinct and comprehensive search queries that effectively help users find the specified item.
Focus on incorporating key features, standout aspects, brand, and practical benefits into each query to enhance search accuracy.
Emphasize the unique attributes that differentiate the item from similar products.
Each query should be concise, factual, and separated by line breaks.
<|eot_id|>
<|start_header_id|>user<|end_header_id|>
### Task:
Analyze the provided item metadata and user reviews to generate 10 detailed and objective search queries for the item.
Your goal is to create queries that highlight the item's key features, benefits, and unique aspects based on its description and user feedback.

### Input:
- **Item Title**: Last Night on Earth: Growing Hunger Expansion
- **Brand**: Flying Frog Productions
- **Description**: The Growing Hunger Expansion introduces new game mechanics and three exciting new Scenarios to challenge players as well as a two player mini-game. Take control of four new Heroes, each with a highly-detailed plastic miniature as well as seven new Red Zombies for use as Plague Carriers, Grave Dead, or to increase the Zombie Horde. New modular game board sections expand the town and feature unique buildings such as the Supermarket, Library, and Antique Shop. New game cards give Zombies a chance to steal weapons from the Heroes and add powerful Double-Handed weapons to the Heroes arsenal, such as Garden Shears and the Fence Post. Also included are two new full color, die-cut counter sheets adding Free Search Markers for the Heroes as well as many more fun ideas to the Last Night on Earth toolbox for limitless use with official web content or creating your own new Scenarios.
-**User Reviews**: 
1. Four Stars Great games
2. Last Night on Earth is the best! Last Night on Earth: Growing Hunger is an expansion for the Last Night on Earth game. I love it because it is so much fun! My friends and I get together once a month for game night and this is one we always pull out. This expansion add heroes, props, locations and scenarios to the main game. You do need the main Last Night on Earth board game for this add on to help you in any way. If you would like to know more about the main game there is a video on you tube from Wil Wheaton on his tabletop gaming blog style show called "TableTop" they played last night on earth https://www.youtube.com/watch?v=UhLU2-BuhMIIts a great explanation of the game and how it works! However I took it to the next level and I hand painted my hero figures. I will post the images.
3. Good expansion It's a good expansion for a good game. I don't know what else to write so I'm finishing with filler.
4. Great expansion to a great game A worthy expansion to the original game, adding new characters, cards, scenarios, and lots and lots of extra components for building your own scenarios and adventures.  For me, thats the big difference between owning just the game and this expansion:  whether or not you intend to make your own scenarios or just play the ones that come with the game.  If you are one of those "game tinkerers" like me, then you will love this expansion.
5. A good addition to a great game. I'll keep it pretty short. If you're considering the expansion, I would hope you already own Last Night on Earth. I think LNOE is a great game that offers suspense, teamwork, competition, and usually a lot of laughs. This expansion does not change much. What it does do is add some more locations (in the form of some new L-shaped boards), some new characters, new hero and zombie cards, and a few new scenarios. On top of that, it comes with new game pieces that are used with the free scenarios available on the games' official website or that can be used to design your own scenarios.In short, this expansion does not reinvent the wheel. What it will do, however, is help expand and reinvigorate an already great game. It's a little pricey, but if you like LNOE, it's a good investment.
6. great game add on The is a great game expansion to the game last night on earth, its very fun and different i highly recommend this add on. love the new game pieces


### Requirements:
- Generate exactly 10 distinct search queries, each on a **separate line**.
- Incorporate key metadata and review insights to create effective, descriptive queries.
- Highlight the item's purpose, standout features, brand, and practical benefits in each query.
- Emphasize the **unique attributes that differentiate the item from other similar products**.
- Avoid redundant details and ensure each query is unique and precise.

### Response:
<|eot_id|><|start_header_id|>assistant<|end_header_id|>
\end{lstlisting}

\subsection{Item Query Sample}

\paragraph{Toys}
\:

\begin{lstlisting}[language=json]
1. **"Radica 20Q AI Game for Kids and Adults: Color Options and Educational Value"**
2. **"20Q Toy with Smart Features: How Does it Guess What You're Thinking?"**
3. **"Best Party Game for Adults and Kids: 20Q Review and Comparison"**
4. **"Portable and Durable 20Q Game for Travel and Classroom Use"**
5. **"Radica 20Q AI Game: How Accurate is it in Guessing Words and Concepts?"**
6. **"Fun and Educational Toy for Kids: 20Q Review and Gift Ideas"**
7. **"20Q Game for Large Groups: Conversation Starter and Icebreaker"**
8. **"Best Gift for Kids and Adults: 20Q Artificial Intelligence Game with Reviews"**
9. **"20Q Toy with Interactive Features: How Does it Engage Kids and Adults?"**
10. **"Radica 20Q Game: Unique Features and Benefits for Home and Classroom Use"**
\end{lstlisting}

\paragraph{Beauty}
\:

\begin{lstlisting}[language=json]
1. **"Fashionwu 10 Pair Long Black False Eyelashes Makeup for Everyday Use, Natural Look with 1cm-1.5cm Length and 3cm Width"**
2. **"Long Black False Eyelashes for Beginners, 10 Pairs with Curved Shape, Perfect for Enhancing Natural Lashes"**
3. **"Fashionwu Black False Eyelashes for Cosplay, 10 Pairs with Thick and Sturdy Design, Great for Drag Queens and Performers"**
4. **"Affordable 10 Pair Long Black False Eyelashes, Good for Special Occasions and Everyday Wear, Natural and Sassy Look"**
5. **"Fashionwu Black False Eyelashes with Glue Strip, 10 Pairs for Long-Lasting Wear, Perfect for Lash Enthusiasts"**
6. **"Long Black False Eyelashes for Beginners, 10 Pairs with Thicker Band and Longer Length, Great for Volume and Drama"**
7. **"Fashionwu Black False Eyelashes for Stage Performers, 10 Pairs with Sturdy Design and Long-Lasting Adhesive"**
8. **"Natural and Sassy Black False Eyelashes, 10 Pairs with Curved Shape and Thicker Band, Perfect for Everyday Wear"**
9. **"Fashionwu Black False Eyelashes for Special Occasions, 10 Pairs with Glitter and Sparkle, Great for Cosplay and Parties"**
10. **"Long Black False Eyelashes for Lash Enthusiasts, 10 Pairs with Thicker Band and Longer Length, Perfect for Volume and Drama"**
\end{lstlisting}

\paragraph{Sports}
\:

\begin{lstlisting}[language=json]
1. **Trijicon NS 3Dot Set (GR/GR) T0963 for Glock: Military-Tested, Bright & Tough Night Sights with 12-Year Tritium Life**
2. **Best Trijicon Night Sights for Glock: 3Dot Set with White Rings for Enhanced Daylight Visibility and Shock Protection**
3. **Trijicon NS 3Dot Set (GR/GR) T0963: Proven Reliability and Accuracy for Combat Handguns**
4. **Trijicon Tritium Night Sights for Glock: 3Dot Set with Aluminum Housing and Silicon Rubber Cushions for Low-Light Performance**
5. **Trijicon NS 3Dot Set (GR/GR) T0963: 12-Year Warranty, Military-Grade Night Sights for Glock and Other Handguns**
6. **Trijicon NS 3Dot Set (GR/GR) T0963: Fast Target Acquisition and Superior Daylight Visibility with White Rings**
7. **Trijicon NS 3Dot Set (GR/GR) T0963 for Glock: Shock-Resistant and Durable Tritium Night Sights**
8. **Trijicon Tritium Night Sights for Glock: 3Dot Set with Enhanced Low-Light Performance and Long-Lasting Tritium Life**
9. **Trijicon NS 3Dot Set (GR/GR) T0963: Fast and Accurate Night Sights for Glock Handguns with White Rings and Aluminum Housing**
10. **Trijicon NS 3Dot Set (GR/GR) T0963: Military-Tested, Proven Night Sights for Glock and Other Handguns with 12-Year Warranty**
\end{lstlisting}

\subsection{User Query Generation Prompt}
\subsubsection{Prompt Format}
\begin{lstlisting}[language=json]
<|begin_of_text|><|start_header_id|>system<|end_header_id|>
You are an intelligent assistant designed to analyze a user's purchase history and behavior to generate **10 effective search queries** for predicting the **next items** they are most likely to purchase.
Your task is to evaluate past purchase patterns, item metadata, and related search queries to construct concise and accurate search queries that can be used to find the next recommended items.
Focus on identifying recurring patterns, shifts in preferences, and evolving interests to enhance the relevance of the search queries.
For the most recent item, make sure to include **related queries** that were associated with it to improve search accuracy.
Ensure each query highlights the **unique characteristics of items** and reflects the **user's preferences and interests** for more personalized recommendations.
Ensure each query is clear, specific, and optimized for retrieving relevant items.
<|eot_id|>
<|start_header_id|>user<|end_header_id|>
### Task:
You are an intelligent assistant tasked with generating **10 optimized search queries** to predict the **next items** a user is likely to purchase based on their chronological purchase history, item metadata, and related search queries.

**Purchase History:** A chronological list of items the user has purchased, including item brands, categories, descriptions, associated metadata, and related search queries. For the **most recent item**, related queries are also provided to enhance search relevance.


### Output Format:
Your response should follow this exact format, ensuring:
1. Each search query is presented on a **separate line**.
2. **No newlines or additional formatting** within each query.
3. The queries should be concise, specific, and optimized for accurate item retrieval.


### Requirements:
- Generate **10 precise search queries** based on the user's purchase history, item metadata, and related search queries.
- For the **most recent item**, ensure that **related queries** are incorporated to improve relevance.
- Ensure each query captures key patterns, preferences, and interests derived from the provided data.
- **Highlight the unique characteristics of items** (e.g., special features, distinctive attributes) and reflect the **user's preferences and behavioral trends** in the queries.
- Do **not** include explanations, introductions, or follow-up comments.
- Keep each query **clear, concise, and limited to a single line**.


### Input:
{user_history}
{generated_queries}

### Output:<|eot_id|><|start_header_id|>assistant<|end_header_id|>
\end{lstlisting}

\subsubsection{User History Example}

\paragraph{Toys}
\:

\begin{lstlisting}[language=json]
**Title:** `Melissa &amp; Doug Wooden Take Along Tool Kit (24pc)`
**Brand:** Melissa &amp; Doug
**Categories:** Toys & Games, Dress Up & Pretend Play, Pretend Play, Construction Tools
**User Review:**
Daughter loves to be like daddy My little girl loves to play pretend and "help" her Daddy and Poppa. She loves to build, hammer, and screw nails and bolts along with them. We have had it about a year and the nails are falling apart - it's not quite as well built as most of the Melissa and Doug products we have had in the past. However, we are still very pleased with the product overall. I especially like the fine motor skills it helps to develop.

**Title:** `Melissa &amp; Doug Penguin Plush`
**Brand:** Melissa &amp; Doug
**Categories:** Toys & Games, Stuffed Animals & Plush, Animals & Figures
**User Review:**
One of her favorite stuffed animals My parents bought this for my daughter last Christmas, and she still loves it. It was as tall as she was last year, so she was adorable at Christmas hugging it and lugging it around the house. Now that she is older, she still loves playing with it and it is easier for her to manipulate.One reason they bought the penguin was due to her love of the movie "Happy Feet". One of her favorite things to do is lie on the floor with her penguin while watching the movie. I don't see much educational value other than supporting (or spurring) an interest in penguins.

**Title:** `Melissa &amp; Doug Farm Sound Puzzle`
**Brand:** Melissa &amp; Doug
**Categories:** Toys & Games, Puzzles, Pegged Puzzles
**User Review:**
Another favorite This is another great Melissa and Doug product my parents bought for my daughter. When she was two, she got this for her birthday. It was so cute to see her face the first time she lifted one of the pieces and heard the animal sound. Personally, we thought the sound quality was fine.The wooden construction provides plenty of durability, and she did not tear it up. We were able to pass it on to her younger cousin. The sound quality was deteriorating slightly by that time, but it was still relatively easy to hear. We had it for about one year before passing it to her cousin.

This is the most recently purchased product:
**Title:** `Melissa &amp; Doug Flip to Win Memory Game`
**Brand:** Melissa &amp; Doug
**Categories:** Toys & Games, Games, Board Games
**User Review:**
Another great product I bought this last year for my now four-year-old daughter. We have always been pleased with the Melissa and Doug product line, and this one does not disappoint. I thought the cardboard "sheets" that you interchange would not last, but they have held up surprisingly well. We only have one that is a bit dog-eared, and that's due to my daughter thinking it would taste delicious so into the mouth it went.It's been a great learning tool. For example, the colors let us explore colors far beyond the basic 6 or 8 most learning tools address. This has increased her vocabulary, and it has made learning much more fun.All in all, this is one of my favorite products - and my daughter likes it a ...
\end{lstlisting}

\paragraph{Beauty}
\:

\begin{lstlisting}[language=json]
**Title:** `Now Foods: Tea Tree Oil, 4 oz`
**Brand:** Now Foods
**Categories:** Beauty, Skin Care, Body, Moisturizers, Oils
**User Review:**
For keloid scars. I bought this to treat my keloid scars and well, it didn't work, but I give it a high rating anyway because it does have other benefits and uses. The smell isn't as bad, it smells like wet wood or wet bark.

**Title:** `Dermablend Quick Fix Concealer SPF 30, Ivory`
**Brand:** Dermablend
**Categories:** Beauty, Makeup, Face, Concealers & Neutralizers
**User Review:**
Works well but wrong color. The color described isn't accurate. This turned out to be a more pinkish color and I am tan/yellowish so it did not work well for me. It is really thick though so it works as far as covering blemishes.

**Title:** `Finulite Cellulite Smoothing Massage Mitt`
**Brand:** Finulite
**Categories:** Beauty, Bath & Body, Bathing Accessories, Bath Mitts & Cloths
**User Review:**
Smooth skin. I used this on my thighs with the Finulite cream and in the shower with regular body soap and it made my skin super smooth. It removes dead skin and you can see it on the massage mitt. It's easy to clean as well. It's a tad painful on dry skin but in the shower when your skin is wet, it glides very smoothly and feels really nice.

**Title:** `Xtreme Brite Brightening Gel 1oz.`
**Brand:** Xtreme Brite
**Categories:** Beauty, Hair Care, Styling Products, Creams, Gels & Lotions
**User Review:**
Mixed feelings. I have mixed feelings about this product. When I first started using it, I can see and feel the difference. It actually started to literally peel the dark skin away. There's a bit of a sting but that's how I know it was working. My skin was lighter in just a couple weeks! However, once I've stopped using it when I reached my desired skin color, it went back to being dark. I applied more and suddenly it was no longer working. It didn't peel or sting. I wonder what went wrong or if it has an expiration date or something. Overall, it works but definitely not consistent.

**Title:** `Remington CI95AC/2 Tstudio Salon Collection Pearl Digital Ceramic Curling Wand, 1/2 Inch - 1 Inch`
**Brand:** Remington
**Categories:** Beauty, Hair Care, Styling Tools, Irons, Curling Irons
**User Review:**
My favorite tool! These are great for natural, beachy waves for your hair. The heatproof gloves work wonders as well. It heats up real hot and is super easy to use.

This is the most recently purchased product:
**Title:** `Finulite - The End to Cellulite, AM/PM Cellulite Cream (2 - 4 oz tubes)`
**Brand:** Unknown
**Categories:** Beauty, Skin Care, Body, Moisturizers, Creams
**User Review:**
I guess it's my fault. The main thing with these creams is you have to be diligent and consistent, which I wasn't. It is a PAIN in the butt to do it every night and every morning, scrubbing with the somewhat painful glove I bought with it. It did make my skin super smooth, but cellulite is a bitch to get rid of even with exercise (lots of thin people I know have cellulite, so it's definitely not a fat people thing). Long story short, I got tired of the routine and gave up.

\end{lstlisting}

\paragraph{Sports}
\:

\begin{lstlisting}[language=json]
**Title:** `Tone Fitness Cement Filled Kettlebell Set - 30 lbs.`
**Brand:** Tone Fitness
**Categories:** Sports & Outdoors, Exercise & Fitness, Strength Training Equipment, Kettlebells
**User Review:**
EXCELLENT Prouct EXCELLENT kettlebell's for both men and women. Vinyl is very smooth and easy on the hands. I would highly recommend.

**Title:** `Blackburn AirTower 2 Bicycle Pump, Silver`
**Brand:** Unknown
**Categories:** Sports & Outdoors, Cycling, Accessories, Bike Pumps, Floor Pumps
**User Review:**
Dy-no-mite This think can fill an auto tire in minutes. The best pump I have ever had. Highly recommend it. Not cheap thin metal. This is good quality thick steel.

**Title:** `Kimber Pepperblaster 2 Red, One Size`
**Brand:** Kimber
**Categories:** Sports & Outdoors, Outdoor Gear, Camping & Hiking, Personal Care
**User Review:**
Welll now... This shoots two shots of Pepper spray. That is it. You throw it away. It is NOT reloadable. It is the same excellent Kimber quality as their weapons are... But how do I know where to aim if I can't practice shoot it ??? I would not recommend this to anyone.

**Title:** `Ultimate Arms Gear Tactical 4 Reticle Red Dot Open Reflex Sight with Weaver-Picatinny Rail Mount`
**Brand:** Ultimate Arms Gear
**Categories:** Sports & Outdoors, Hunting & Fishing, Hunting, Hunting Optics, Gun Scopes, Rifle Scopes
**User Review:**
Excellent Fits my gun great. Easy to mount. Very large so the image is clear. No reason you should ever miss a target with this site. GREAT quality.

**Title:** `Camelbak Podium Big Chill 25 oz Bottle`
**Brand:** CamelBak
**Categories:** Sports & Outdoors, Accessories, Sports Water Bottles
**User Review:**
Ahhh.... Hard to get enough water out of this bottle. Doesn't keep anything cold. Very thin material. I am worried after being out in the sun, how long before it breaks.

**Title:** `Kimber Pepperblaster Ii Holster`
**Brand:** Meprolight
**Categories:** Sports & Outdoors, Paintball & Airsoft, Airsoft, Holsters
**User Review:**
As I said in the other feedback If you think you need the Kimber Pepperblaster then this is a great holster to carry it in. Now that I found out you can't reload you Pepperblaster, I see no reason for either the blaster or the holster. How do you practice ...???

**Title:** `5 LED Bicycle Rear Tail Red Bike Torch Laser Beam Lamp Light`
**Brand:** 
**Categories:** Sports & Outdoors, Cycling, Lights & Reflectors, Taillights
**User Review:**
Great lite Very bright. I would recommend this lite to anyone. I put one on my wife's bike. I love this light.

This is the most recently purchased product:
**Title:** `Body Solid MA307N Nylon Head Harness`
**Brand:** Body Solid
**Categories:** Sports & Outdoors, Exercise & Fitness, Exercise Machine Accessories, Exercise Machine Attachments
**User Review:**
Exactly what I wanted My friend runs a boxing school for urban kids. These kids work hard at staying out of trouble, doing their school work and being able to learn to box. This harness works perfect. It is of highest quality.

\end{lstlisting}

\subsection{User Query Sample}
\subsubsection{Toys}
\begin{lstlisting}[language=json]
1. "LeapFrog TAG Reading System Purple 32MB pen with phonetic sound and games for 4-year-olds, educational toy for learning to read and spell"
2. "Best LeapFrog reading system for preschoolers with built-in games and activities, suitable for travel and home use"
3. "LeapFrog TAG Reading System Purple review, educational toy for kids to learn alphabet, spelling, and reading skills with phonetic sound and games"
4. "LeapFrog TAG Reading System 32MB pen with LeapFrog Connect software for downloading and managing content, suitable for children aged 3-6 years"
5. "LeapFrog TAG Reading System Purple with headphones, compact and portable learning tool for kids to learn to read and spell with fun games and activities"
6. "LeapFrog TAG Reading System Purple review, educational toy for kids to learn phonics, sight words, and reading skills with LeapFrog Connect software and games"
7. "LeapFrog TAG Reading System Purple 32MB pen with built-in games and activities, suitable for kids to learn to read and spell with phonetic sound and fun stories"
8. "LeapFrog TAG Reading System Purple with LeapFrog Connect software for downloading and managing content, suitable for children aged 3-6 years, educational toy for learning to read and spell"
9. "LeapFrog TAG Reading System Purple review, educational toy for kids to learn alphabet, spelling, and reading skills with phonetic sound, games, and activities"
10. "LeapFrog TAG Reading System Purple 32MB pen with LeapFrog Connect software and games, suitable for kids to learn to read and spell with fun stories and activities, compact and portable"
\end{lstlisting}

\subsubsection{Beauty}
\begin{lstlisting}[language=json]
1. "Neutrogena Microdermabrasion System with glycerin and ultra-fine crystals for smooth, luminous skin and anti-aging benefits, clinically proven for visible results in one use."
2. "Best affordable microdermabrasion system for acne-prone skin, gentle exfoliation, and firming, with 12 pre-dosed puffs and AA batteries included."
3. "Neutrogena Microdermabrasion System with massaging micro-vibrations for improved skin texture, radiance, and fine line reduction, suitable for sensitive skin types."
4. "Microdermabrasion system for at-home use, with dermatologist-recommended Neutrogena brand, proven to deliver smoother, more luminous skin in just one use, and affordable price point."
5. "Exfoliating microdermabrasion system with gentle, pre-dosed puffs and soothing glycerin, ideal for daily use, with visible results in just one treatment, and suitable for all skin types."
6. "Neutrogena Microdermabrasion System with anti-aging benefits, firming, and skin brightening, with a unique combination of exfoliating crystals and micro-vibrations, and easy to use at home."
7. "Best microdermabrasion system for oily skin, with a gentle, non-irritating formula, and visible results in just one use, with a affordable price and convenient packaging."
8. "Neutrogena Microdermabrasion System with a dermatologist-recommended formula, proven to deliver smoother, more radiant skin, and suitable for daily use, with a unique combination of exfoliation and micro-vibrations."
9. "Microdermabrasion system for anti-aging, firming, and skin brightening, with a gentle, pre-dosed puff system, and visible results in just one treatment, and suitable for sensitive skin types."
10. "Neutrogena Microdermabrasion System with a clinically proven formula, delivering visible results in just one use, and a affordable price point, with a unique combination of exfoliation and micro-vibrations for smoother, more luminous skin."
\end{lstlisting}

\subsubsection{Sports}
\begin{lstlisting}[language=json]
1. "ProSource Heavy-Duty Easy Gym Doorway Chin-Up/Pull-Up Bar with 300lb weight capacity and multi-position design for home workout"
2. "Best doorway pull-up bar for thin walls and easy installation with ProSource Comfort Grip technology"
3. "Heavy-duty pull-up bar for home gym with adjustable grip and sturdy construction for 24-32 inch doorways"
4. "ProSource Easy Gym Doorway Chin-Up/Pull-Up Bar with wall-mounting option and 5-star durability rating"
5. "Inexpensive and easy-to-assemble pull-up bar for home workout with 300lb weight capacity and multi-functional design"
6. "Best pull-up bar for doorways with raised molding and minimal wall damage with ProSource brand guarantee"
7. "Heavy-duty doorway pull-up bar with adjustable grip and comfortable grip technology for home gym workouts"
8. "ProSource Heavy-Duty Easy Gym Doorway Chin-Up/Pull-Up Bar with 300lb weight capacity and easy installation for home fitness"
9. "Best pull-up bar for doorways with multi-position design and sturdy construction for home workout and exercise"
10. "ProSource Easy Gym Doorway Chin-Up/Pull-Up Bar with 300lb weight capacity and wall-mounting option for home gym and fitness enthusiasts"
\end{lstlisting}

\newpage

\end{document}